\documentclass[10pt]{aastex}
\usepackage{bm}
\usepackage{lscape}
\usepackage[dvips]{epsfig}
\usepackage{amsmath}
\usepackage{amssymb}
\usepackage{amsthm}
\usepackage{array}
\usepackage{multirow}
\usepackage{graphicx}
\usepackage{amsfonts}
\usepackage{eucal}
\begin{document}

\newcommand{\oq}{\textquotedblleft}
\newcommand{\cq}{\textquotedblright}
\newcommand{\cqb}{{\textquotedblright}~}
\newcommand{\ms}{$M_{\odot}$}
\newcommand{\msb}{$M_{\odot}$~}
\newcommand{\ct}{$^{13}$C}
\newcommand{\ctb}{$^{13}$C~}

\graphicspath{{figures/}}

\title{Isotope Anomalies in the Fe-group Elements in Meteorites and Connections to Nucleosynthesis in AGB Stars}

\author{G.J. Wasserburg}
\affil{Lunatic Asylum, Geological \& Planetary Sciences, California Institute of Technology, Pasadena, CA 91125, USA; gjw@gps.caltech.edu}

\author{O. Trippella, M. Busso}
\affil{Department of Physics \& Geology, University of Perugia, and INFN, Section of Perugia, via A. Pascoli, Perugia 06123, Italy; oscar.trippella@fisica.unipg.it; maurizio.busso@fisica.unipg.it}

\begin{abstract}
We study the effects of neutron captures in AGB stars on \oq Fe-group\cqb elements, with an emphasis on Cr, Fe, and Ni. These elements show anomalies in $^{54}$Cr, $^{58}$Fe, and $^{64}$Ni in solar-system materials, which are commonly attributed to  SNe. However, as large fractions of the interstellar medium (ISM) were reprocessed in AGB stars, these elements were reprocessed, too. We calculate the effects of such reprocessing on Cr, Fe, and Ni through 1.5\msb and 3\msb AGB models, adopting solar and 1/3 solar metallicities. All cases produce excesses of $^{54}$Cr, $^{58}$Fe, and $^{64}$Ni, while the other isotopes are little altered; hence, the observations may be explained by AGB processing. The results are robust and not dependent on the detailed initial isotopic composition. Consequences for other \oq Fe group\cqb elements are then explored. They include $^{50}$Ti excesses, and some production of $^{46,47,49}$Ti. In many circumstellar condensates, Ti quantitatively reflects these effects of AGB neutron captures. Scatter in the data results from small variations (granularity) in the isotopic composition of the local ISM. For Si, the main effects are instead due to variations in the local ISM from different SNe sources. The problem of Ca is discussed, particularly with regard to $^{48}$Ca. The measured data are usually represented assuming terrestrial values for $^{42}$Ca/$^{44}$Ca. Materials processed in AGB stars or sources with variable initial $^{42}$Ca/$^{44}$Ca ratios can give apparent $^{48}$Ca excesses/deficiencies, attributed to SNe. The broader issue of Galactic Chemical Evolution is also discussed in view of the isotopic granularity in the ISM. \end{abstract}


\section{Introduction}
The purpose of this study is to investigate the effects of slow neutron captures in AGB stars on isotopes of the \oq Fe group\cqb elements. In particular Ti, Cr, Fe, Ni, and Zn are considered to be members of the \oq iron group\cq. These nuclei are predominantly the product of supernovae (SNe), both core collapse and SNeIa \citep[see e.g.][and references therein]{mo14,cl13, lan13, hil00}. It has been known for over three decades that small ($\lesssim$ 1\%) variations are found in the isotopic compositions of Ti, Cr, Fe, Ni, and Zn in macroscopic samples of meteorites: see e.g. \citet{ir90,i+00} and the review by \citet{Birck}. In particular, these isotopic effects are common in calcium-aluminum-rich inclusions (hereafter CAIs) in meteorites, but are also wide spread at lower levels in \oq bulk\cqb samples of different groups of meteorites. The observed effects in Cr, Fe, and Ni typically show enrichments in the heaviest isotopes $^{54}$Cr, $^{58}$Fe, and $^{64}$Ni, which are a small fraction of the abundance of the respective element ($\sim$ 2.3\% for $^{54}$Cr, $\sim$ 0.28\% for $^{58}$Fe, $\sim$ 0.9\% for $^{64}$Ni). In addition, the isotopic patterns observed in Ti, Ca, and Si are distinct and complex. Only one datum exists for Zn. As the Fe group elements are associated with Supernovae (SNe), it has been natural that interpretations of these effects in CAIs and \oq bulk\cqb meteorites have focused on nucleosynthetic processes in them. This has not often led to any firm explanation of the observed isotopic anomalies or to clear predictions. The first study that considered the possibility of $s$-process effects on Fe-group nuclei was by \citet{the}, who explored the occurrence of $s$-processing in massive stars after revisions in the rate for the $^{22}$Ne($\alpha$,n)$^{25}$Mg reaction and provided some possible insights. 

The $s$-processed material is, in the interstellar medium (hereafter ISM), predominantly from low mass AGB stars. We expect that the effect of $s$-processing on Cr, Fe, and Ni (originally synthesized in SNe) is to produce the heavier isotopes of these elements, $^{54}$Cr, $^{58}$Fe, and $^{64}$Ni, in over-abundance; this can be seen by considering that the pure $s$-process abundance pattern is, in general,  smoothly correlated with the neutron-capture cross sections, yielding processed material enriched in nuclides heavier than the seed nuclei. As only a very small fraction of matter in the star is subjected to $s$-processing, it is not obvious just how large the consequences will be in the envelope of an AGB star, or will be provided to the ISM. There are thus three distinct issues: i) What are the $s$-process effects on isotopic abundance patterns in Cr, Fe, and Ni in an AGB envelope? ii) How significant are these effects? iii) What are the effects to be expected on the nuclei of lower atomic number than iron (e.g. Ti, Si, and Ca)?

We will show that there are substantial consequences on the abundances of $^{54}$Cr, $^{58}$Fe, and $^{64}$Ni produced in AGB stars and that these may provide an explanation of the effects observed in many macroscopic meteorite samples. We will try to relate this to data in presolar dust grains. The shifts found in other Fe-group nuclei subject to the $s$-process will be shown to be more complex. 

In the different classes of data that are available we must consider both presolar grains, each of which come from a particular stellar source, and macroscopic samples of solar system material, that come from blends from different stellar sources.

The macroscopic samples of very early solar system matter derive from mixing, heating, melting from various precursor materials and exhibit relatively small isotopic shifts ($<$ 1\%). They are blends of very local parcels from unknown sources, but mixed so that they are rather close to the average isotopic admixture. In contrast to such macroscopic samples, grains formed as circumstellar condensates (subsequently called CIRCONs) and found in meteorites may exhibit very large ($>$ 50\% or more) isotopic abundance shifts \citep[see e.g.][for an in-depth review]{zinner14}. We will try to relate the results on Cr, Fe, and Ni to data on presolar dust grains. A CIRCON is the direct product of stellar nucleosynthesis in a single star. The initial material from the ISM that is incorporated in a forming star may be variable, but mixing provides an initially homogeneous isotopic pattern in that star, which is subject to processing in the subsequent stellar evolution before the formation of a grain. The effects produced will depend on the particular stage in the evolution and on the initial isotopic composition of the star. Results on CIRCONs often serve as clues to still unrecognized stellar processes. There are often difficulties in explaining nucleosynthesis effects in a circumstellar dust particle from a single star resulting from incomplete knowledge of many nuclear and dynamical processes. For an individual CIRCON grain, the isotopic composition then depends on the initial abundances (not known) of the associated star and on the details of the subsequent stellar evolution which we claim to understand. For stellar sources that have, at each time, a well mixed envelope, the problem is simplified. In this study, we will focus on those CIRCONs that are inferred to come from AGB stars and not SNe. For SNe sources, it is possible to find different production zones in the models, with different yields, and presume that these can be sampled ad-hoc to produce the isotopic pattern and the carbon compounds observed. Grains that are plausibly attributed to SNe sources are not considered here. The problem of forming C-rich phases in standard SNe sources with a high O/C ratio is not really addressed. The case of AGB stars is much more restricted, as their envelopes are well mixed.

As there must be a spread in the isotopic abundances in the ISM (both in time and space), different star formation sites will have some variability in their \oq bulk\cqb isotopic patterns. Thus, even if we know the exact evolution of an AGB star in terms of its mass and metallicity, it is not possible to infer the resulting isotopic shifts. However, the isotopic data on CIRCONs give good insight into the nature of the nuclear effects and, often, a rather clear identification of the possible stellar site. We will therefore use some of the data on such circumstellar condensates that are associated with AGB stars in conjunction with processed Solar System samples to explore the implications of $s$-processing models of Fe group elements. This requires an interplay of data from Solar System processed material and the data on CIRCONs. We will also use this approach to infer aspects of the isotopic variability in the different samples of ISM which the CIRCONs represent and to explore the general isotopic anomalies found in solar-system processed material. In particular, it will be shown that the Ti data on CIRCONs demonstrate that they show the results of neutron captures and that the complex patterns observed are simply a reflection of AGB processing on material with some small, but substantial variability in isotopic abundances in the original stellar sources. The questions of $s$-processing of Si and Ca and of variability within the ISM will also be discussed.

Unlike circumstellar condensates, the solar system was formed out of a complex assembly of stellar debris, from a wide variety of sources, operative at greatly different times, which were not completely mixed. While there is clear isotopic evidence for the contributions from certain precursor sources to individual macroscopic objects processed in the solar system, it is not possible to explain the observed isotopic variability in macroscopic samples by a single mechanism or a single star. What we observe are changes in the isotopic composition of an element relative to some terrestrial standard. This means that large isotopic shifts may be more easily interpretable than very small effects. The isotopic ratios of the \oq bulk\cqb solar system is, of course, not exactly known. Positive variations in isotopic abundances may be easily interpreted as \oq additions\cqb of an isotope to the solar reservoir; but the presence of negative variations clearly show that we are dealing with isotopic admixtures that may be different for many isotopes, as has been well recognized. 

We first present results for Cr, Fe, and Ni for AGB sources that have initial isotopic abundances which are solar. Then, we will explore the characteristics of $s$-processing when the source is not solar and show that there are simple rules which permit one to estimate relative isotopic shifts for an element independent of the mass or metallicity of the particular AGB star. This leads to a general rule for stars which do not have an initial inventory with solar isotopic abundances. The discussion will extend to the broader issue of whether the observational evidence supports processing in low mass AGB stars.

\section{The Calculations}
\subsection{The stellar and nucleosynthesis models}
The $s$-processing in AGB stars takes place on a small parcel of matter that is removed from the stellar envelope and is exposed to neutrons. This processed material is then mixed back into the mass of the envelope present at that event \citep{b+99,k+11}. As the envelope looses mass, the dilution of the processed material continuously changes during the thermally-pulsing (TP) phase associated with the $s$-process. Also the seed nuclei of the star are to some extent altered by the processing required to produce the bulk of the $s$-nuclei. To produce the \oq solar\cqb $s$-inventory requires $\sim$ 10$^{-2}$ of all the seeds to be consumed. This will result in a small, but significant depletion of the seed nuclei in the composition of the envelope. The isotopic shifts in the envelope will be small for each cycle of $s$-processing, but they will increase in time as the envelope is greatly reduced in mass. While usual discussions of the $s$-process refer to the Fe group elements as the seeds, in fact there is also a major involvement of lower mass species, particularly if one considers elements like Ca, Ti, and Si, whose final concentration is due both to their initial inventory and to the modifications induced by neutron captures on lighter, abundant elements.

The calculations presented here assumed that the initial inventory of all heavy elements was solar, with the solar isotopic ratios taken from \citet{Lod}, but with the abundances of \oq heavy elements\cqb relative to H and He variable as an estimate of the net metallicity of the system \citep[see e.g. the same procedure adopted in][]{was06}. The net number of isotopes ($N_i$) in the envelope for all species is calculated at the end of each pulse-interpulse episode. This includes the $s$-processed material just produced and dredged up and the material already present in the envelope from previous episodes and not lost. Calculations were carried out through the full evolution of model AGB stars, including all stages in which any significant neutron fluence occurs. These were done up to the end of the AGB phase. The operation of the main neutron source $^{13}$C($\alpha$,n)$^{16}$O was followed taking the size of the $^{13}$C reservoir (or {\it pocket}) from \citet{t+14}. A second burst of neutrons occurs from the $^{22}$Ne($\alpha$,n)$^{25}$Mg reaction in the thermal pulses; its relevance varies with the model, as the rate of neutron production is sensitive to temperature. We report on stars ranging in mass from 1.5 to 3\ms. The 3\msb case always evolves to an envelope with C/O $>$ 1, while the 1.5\msb case becomes C-rich only at relatively low metallicity or after very advanced dredge-up stages. In the 3\msb case, the neutrons from the $^{22}$Ne source are important and produced at a rather high neutron density, while in the lower mass models the $^{13}$C source always dominates. The $^{13}$C pocket used by \citet{t+14} was formed adopting an exponential profile for the proton penetration down to $6\times10^{-3}$ \msb (depth in mass). In this treatment most of the available \ctb is within about $4\times10^{-3}$ \ms, which produces the galactic enrichment of $s$-elements described in \citet{mai12}. The reactions considered in the region of Fe-peak nuclei are shown in Figure \ref{network}, where we also give approximate 30keV Maxwellian-averaged values for cross sections \citep[in mb, KADONIS database][]{k+11} and laboratory half-lives of crucial nuclei of interest \citep{TAK87}. In the detailed calculations used, cross sections variations are considerable, and generally more complex than a simple $1/v$ scaling; also decay rates may change.

The mass loss rates were taken from the new picture of low mass star winds deriving from the recent works by \citet{g+06} and \citet{gc13}, where the model AGB luminosities could be reconciled with observations. Cases for different metallicities $Z$ were studied; here the notations used are $Z=Z_{\odot}$ for a total metallicity of $Z$ = 0.0154 \citep{Lod}; and $Z=Z_{\odot}/3$. At low [Fe/H] values, all heavy elements are scaled downward in same way as iron. The nucleosynthesis code was that used by \citet{t+14}; however, the calculations presented here are somewhat different from those reported in \citet{t+14} as they now include explicitly $^{50}$Cr (the full computer output is available from the authors on request).

For each element, we have chosen an index nucleus ($k$) which is of high abundance and critical to any flow induced by n captures in producing isotopes of higher mass of that element. For chromium, iron and nickel the index isotopes were $^{52}$Cr, $^{56}$Fe, and $^{58}$Ni. Note that some isotopes are bypassed or purely destroyed in the neutron capture chain (e.g. $^{54}$Fe and $^{50}$Cr). The production of $^{64}$Ni depends on the complex branch at $^{64}$Cu. The general reaction network is shown in Figure \ref{network}.

\subsection{Cr, Fe, and Ni in model AGB envelopes}
The enrichment factors ($E^{\odot}_{i,k}$) in isotopic abundances for Cr, Fe, and Ni were calculated for the envelopes of 1.5 and 3 \msb model stars for $Z = Z_{\odot}$ and $Z = Z_{\odot}/3$ after a different number of pulses; they are shown in Figure \ref{over1} and Figure \ref{over2}. These calculations are for solar initial isotopic abundances. We define the plotted enrichment factor in the envelope at some particular stage ($S$), for the ratio of the number of nuclides {\it i} to that of the index nuclei $k$, relative to the assumed initial state (${\odot}$) in the star, as:
$$
E^{\odot}_{i,k}=\frac{\left(\dfrac{N_i}{N_k}\right)^S}{\left(\dfrac{N_i}{N_k}\right)^{\odot}} -1. \eqno(1)
$$
We have also included in Figures \ref{over1} and \ref{over2} the results taken from the calculations by \citet{c+11}. In the report by \citet{c+11} the focus was on the relative yield of heavy to light $s$-process nuclei. However, the on-line data (available from the FRUITY site at: http://fruity.oa-teramo.inaf.it/) represent a full enough calculation of important nuclei in the region of interest \citep{cpo14}. They can thus be applied to the problem of Cr, Fe, and Ni. It is evident that both models produced marked excess of $^{54}$Cr, $^{58}$Fe, and $^{64}$Ni. This general inference thus appears well established. There are clear differences found between our models and FRUITY data as can be seen by careful inspection of Figure \ref{over1} and \ref{over2}. We note that the enrichments of $^{54}$Cr and $^{58}$Fe are much lower than that found in the FRUITY data for a 1.5\msb source. There is larger production of $^{61}$Ni in the FRUITY models in the normalized ratios shown in Figures \ref{over1} and \ref{over2}, but the major excesses are always at $^{64}$Ni. 

These issues will be noted in the text. For reasons of self consistency, the general discussion and conclusions presented here will use the results following \citet{t+14} who present higher neutron fluences and slightly higher temperatures than in the FRUITY models. We believe that the specific issues with these AGB models will require further detailed investigation. The differences found between our models and the FRUITY data are predominantly due to the different size of the \ct-pocket used in the calculations (our pocket is about four times larger than for FRUITY), to the different solar metallicity assumed and to the non-linear effects induced by them. Inspection of Figures \ref{over1} and \ref{over2} shows in any case that the results are generally in agreement and there is no qualitative conflict between these independent works. In particular only the FRUITY data set and our results specifically treat the production of $^{54}$Cr, $^{58}$Fe, and $^{64}$Ni, including the effects from the presence of $^{50}$Cr and of $^{54}$Fe, which are not relevant for computing the production of much heavier $s$-nuclei. We note that all other calculations available in the literature and on-line either do not have envelope abundances for Cr, Fe, and Ni, or are not adequate to permit inferences about yields in this atomic mass region. 

We also notice that a full and detailed comparison with FRUITY is difficult to make, due to different assumptions. As an example, if we look at Ti isotopes, FRUITY outputs give a low production for $^{47}$Ti and enhancements for $^{49}$Ti and $^{50}$Ti at high levels (see Figure \ref{tifr}). This is strange, considering the small pocket used in the code. However, FRUITY adopted a total solar metallicity of $Z_{\odot}$ = 0.014, as in \citet{Asp}, against $Z_{\odot}$ = 0.0154, adopted by us and by \citet{Lod}. This represents a 10\% shift in the solar $Z$. It is possible that these 
small differences in the inputs cause significant  variations in the outputs. This remains to be clarified. 

We consider that the results presented here are a reasonable effort at exploring effects of $s$-processing on the Cr, Fe, and Ni isotopes. It can be seen that the enrichment patterns all show clear enhancements in $^{54}$Cr, $^{58}$Fe, and $^{64}$Ni and very small effects for the other Cr, Fe, and Ni isotopes. For later pulses the $E^{\odot}_{54,52}$, $E^{\odot}_{58,56}$, $E^{\odot}_{64,58}$ are quite large. It follows that AGB processing will produce significant shifts in the isotopic abundances of these three elements which resemble some of the observed patterns. The magnitude of the enrichment factors is dependent on the size of the \ct-pocket, the stellar mass and the stage of evolution of the star. Therefore the pattern will only be approximately fixed. A more detailed discussion for several Fe group elements will be given later.


\subsection{General representation of isotopic results}
\subsubsection{General isotope pattern from all $s$-process}
To gain some insight into the possible isotopic abundance patterns resulting from the updated neutron capture scenario outlined above, we first assume, for simplicity, that the composition of the envelope in a star was initially solar, and that it evolves with the addition of pure $s$-process material. The result for the envelope is simply the sum of what was brought up from the neutron capture zones ($n_i$ or $n_k$) plus the initial inventory ($N_i^{\odot}$ or $N_k^{\odot}$). We then have:
$$
N_i=N_i^{\odot}+n_i = \left(\frac{N_i^{\odot}}{N_k^{\odot}}\right)N_k^{\odot} + \left(\frac{n_i}{n_k}\right)n_k \eqno(2)
$$
$$
\frac{N_i}{N_k} = \left(\frac{N_i^{\odot}}{N_k^{\odot}}\right)f_k + \left(\frac{n_i}{n_k}\right)\left(1-f_k\right) \eqno (3)
$$
where $f_k$ is the fraction of the original abundance of the isotope $k$ remaining in the material processed by a pulse. The fractional shift of the isotopic ratios of the mixture relative to the initial state ({\it solar}, or \oq${\odot}$\cq) is thus:
$$
E^{{\odot},mix}_{i,k}=\left(1-f_k\right)\left[\left(\frac{n_i}{n_k}\right)/\left(\frac{N_i^{\odot}}{N_k^{\odot}}\right)-1\right] \eqno (4)
$$
and for two pairs of isotopes: 
$$
\dfrac{E^{\odot,mix}_{i,k}}{E^{\odot,mix}_{j,k}} = \left[\left(\frac{n_i}{n_k}\right)/\left(\frac{N_i^{\odot}}{N_k^{\odot}}\right)-1\right] / \left[\left(\frac{n_j}{n_k}\right)/\left(\frac{N_j^{\odot}}{N_k^{\odot}}\right)-1\right] \eqno(5)
$$
If $n_i/n_k$ is fixed, this implies that the ratio of the enrichment factors for two isotopes is constant if production ratios are constant through all stages of a stellar evolution. Note that $f_k$ is not the same for different elements.

Inspection of equation (4) shows that the enrichment of a parcel of matter, once diluted with some original envelope material, will have an isotopic pattern that is essentially fixed and independent of the amount of diluting material, while the magnitude of the isotopic shifts will depend on $f_k$, the fraction of the diluting  material. The critical parameter determining the pattern can be indicated as $Q^{\odot}_{i,k}$, where:
$$
Q^{\odot}_{i,k} = \dfrac{E^{{\odot},mix}_{i,k}}{(1-f_k)} = \left[\left(\dfrac{n_i}{n_k}\right) / \left(\dfrac{N_i^{\odot}}{N_k^{\odot}}\right)-1\right]
\eqno(6)
$$
is the enrichment factor in the processed parcel. Insofar as $Q^{\odot}_{i,k}$ is constant for all pulses, then this gives the {\it isotopic abundance pattern produced by pure $s$-processing}. The magnitude of the isotopic shifts in the envelope for a mixture can then be obtained as: 
$$
E^{\odot,mix}_{i,k} = Q^{\odot}_{i,k} (1-f_k) \eqno (7)
$$
To estimate $Q^{\odot}_{i,k}$ we have taken, from our calculations, the outputs for the isotopic abundances in a parcel of matter that was removed from the envelope, subject to neutron exposures, following the procedure described previously, but prior to mixing back into the envelope. This corresponds to exploring the situation in the He-rich layers after a pulse (we arbitrarily chose the last pulse for a 1.5\msb star with solar metallicity).

The curves for $Q^{\odot}_{i,k}$ for the single $s$-processed parcel, {\it assuming a solar initial composition}, are shown in Figure \ref{q_values} and Table \ref{tab1}. In the figure we also show the $Q$ values for a smaller $^{13}$C pocket from the calculations of \citet{t+14} (Case A) for comparison. The extension of this small pocket is $1\times10^{-3}$\msb and it contains $3.1\times10^{-6}$\msb of $^{13}$C. 

\begin{table}[t!!]
\begin{center}
\caption{\label{tab1}Summary of the $Q_{i,k}$ values for a star of 1.5\msb and solar metallicity, as shown by blue-open circles in Figure \ref{q_values}. Ca, Ti, Cr, Fe, Ni, and Zn abundances are calculated using the $^{13}$C pocket described in \citet{t+14}}
\begin{tabular}{ccccccccccccccccc}
\hline
\hline
\multicolumn{2}{c}{Ca} & & \multicolumn{2}{c}{Ti} & & \multicolumn{2}{c}{Cr} & & \multicolumn{2}{c}{Fe} & & \multicolumn{2}{c}{Ni} & & \multicolumn{2}{c}{Zn}\\
\cline{1-2}\cline{4-5}\cline{7-8}\cline{10-11}\cline{13-14}\cline{16-17}
$i$ & $Q_{i,40}$ & & $i$ & $Q_{i,48}$ & & $i$ & $Q_{i,52}$ & & $i$ & $Q_{i,56}$ & & $i$ & $Q_{i,58}$ & & $i$ & $Q_{i,64}$ \\ 
\cline{1-2}\cline{4-5}\cline{7-8}\cline{10-11}\cline{13-14}\cline{16-17}
$40$ & 0.0000 & & $46$ & 3.4887 & & $50$ & -0.565 & & $54$ & -0.379 & & $58$ & 0.0000 & & $64$ & 0.0000 \\
$42$ & 3.4598 & & $47$ & 1.3129 & & $52$ & 0.0000 & & $56$ & 0.0000 & & $60$ & 2.6721 & & $66$ & 5.9473 \\
$43$ & 4.3513 & & $48$ & 0.0000 & & $53$ & 0.4613 & & $57$ & 4.9091 & & $61$ & 19.169 & & $67$ & 9.4356 \\
$44$ & 1.4366 & & $49$ & 9.8874 & & $44$ & 5.4188 & & $58$ & 34.145 & & $62$ & 12.902 & & $68$ & 16.2428 \\
$46$ & 8.9000 & & $50$ & 19.369 & & $  $ &        & & $  $ &        & & $64$ & 175.76 & & $70$ & 0.2361 \\
$48$ & 0.2906 & & $  $ &        & & $  $ &        & & $  $ &        & & $  $ &        & & $  $ &        \\
\hline
\end{tabular}
\end{center}
\end{table}

We note that the $Q$ values for $^{54}$Cr, $^{58}$Fe, and $^{64}$Ni are large as compared to other isotopes of the same elements. They are therefore not sensitive to the initial condition. The $Q$ values are directly dependent on the size of the \ctb pocket, so that this parameter must be taken in full consideration. If we restrict our discussion to a fixed pocket size, then some general rules apply. General patterns will be approximately the same for different pocket sizes, but will be shifted upwards or downwards as can be seen in Figure \ref{q_values} (lower curves).

We now compare the results for $Q_{i,k}$ to the pattern obtained in the full calculations of $E_{i,k}$ in the envelope, for our standard \ctb pocket as given in \citet{t+14}. These enrichment factors are shown in Figure \ref{over} and may be compared with the approximation represented by $Q_{i,k}$ shown in Figure \ref{q_values}. All the results are normalized to the isotope with the largest shift for each element at different stages. This exhibits a comparison of the isotopic abundance pattern for each element as a result of the full calculation. The results for other Fe group elements are also shown in Figure \ref{over}. It is clear that all the $s$-processing, almost independent of stage or metallicity, gives essentially the same isotopic pattern, but with different magnitudes of the enrichment for a fixed \ctb pocket. Further, this pattern is almost indistinguishable from the quantitative pattern approximately obtained by using the $Q^{\odot}_{i,k}$ reference values. We can now explore the dependence of $Q_{i,k}$ on the role of the neutron fluence. 

We used a reference state for $Q$ for the last pulse of a 1.5\ms, $Z = Z_{\odot}$ model AGB. First, note that a key term in the denominator in $Q_{i,k}$ is $n_k$, the number of index nuclei returned to the envelope by the processed parcel. For Fe, the index isotope is also effectively the remaining $^{56}$Fe inventory in the processed parcel. As $^{56}$Fe is not produced from lower mass nuclei, it is only destroyed. Higher neutron fluences cause a decrease in $^{56}$Fe in the returning parcel. Hence, an increase in the fluence would mean a decrease in $Q_{i,56}$. For Ni, the index isotope is $^{58}$Ni: it is mainly destroyed, but also produced by inflow from neutron captures on Fe (which produce more Ni). Thus, there is a much enhanced increase in $^{64}$Ni as compared with $^{58}$Fe. Similar arguments follow for Ti and Cr. In Figure \ref{q_values} we compare our results for a large $^{13}$C pocket \citep[like that used by][]{t+14} with those from a smaller $^{13}$C reservoir of $10^{-3}$ \msb \citep[as discussed e.g. in][]{g+98,was06}. 

The effects of $s$-processing on the parcel considered clearly define the isotopic patterns for all models explored. It appears that the above approximation is robust and reasonably accurate. It is importance to recognize that details for each pulse are not required to describe the resulting isotopic enrichment \textit{pattern}. 

It follows that, for a given \ctb pocket, the $s$-process produces {\it relative} isotopic shifts of an element which are, to a good approximation, independent of the metallicity, model or stage of the particular AGB star, provided the star had initially solar isotopic ratios for the elements of interest. 

\subsubsection{The Relative Magnitudes of the Enrichment Factors $E_{i,j}$}
While the isotopic abundance pattern in the envelope is almost an invariant for all $s$ processing models, the magnitude of the effects for each element is very sensitive to the number of pulses, the efficiency of TDU,the abundance of $^{14}$N in the processed parcel, metallicity and the stellar mass. It is also dependent on the size 
of the \ct-pocket used. In Table 2 we show the relative magnitudes of $E_{i,j}$ at the last pulse for Ti, Cr, Fe, and Ni in models of different mass and metallicity, but with Solar initial isotopic ratios. Only the isotopes with large shifts in patterns are considered here ($^{50}$Ti, $^{58}$Fe, $^{64}$Ni). Data are taken from the inserts in Figure \ref{over}. 

It can be seen that the relative effects for these isotopes with major shifts in 
patterns relative to $^{50}$Ti are very different. $^{54}$Cr is enhanced much less than $^{50}$Ti in all cases. However, $^{58}$Fe is enhanced as much as or more than $^{50}$Ti. For $^{64}$Ni, the enhancement is much greater than for $^{50}$Ti. It appears that this is a general rule. The relative effects to be observed depend on the scaling between different elements. In consideration of variability in the initial isotopic composition, all samples from AGB sources with few pulses will not show significant or recognizable effects. Stars formed at early epochs from low metallicity clouds will show the largest effects.

\subsubsection{Measured Isotopic Shifts and the Initial Composition of a Stellar Source}
It is widely recognized that the initial isotopic compositions for stellar sources which formed from the ISM are variable. To discuss the isotopic effects resulting from AGB processing of initially non-solar isotopic patterns, we must explore the consequences of such variability. The numerical calculations typically assume solar isotopic ratios. In consideration of variability in the initial isotopic composition, all samples from AGB sources with few pulses will not show signiﬁcant or recognizable effects. Only those samples of stars with advanced pulses will show clear signs of $s$-processing. Low metallicity from earlier events will show the largest shifts.

In case of a non-solar initial composition, we define the enrichment factor in the envelope at some particular stage ($S$), for the ratio of the number of nuclides {\it i} to that of the index nuclei $k$, relative to an assumed initial ({\it non-solar}) state \oq $O$\cqb in the star, as:
$$
E^{O}_{i,k}=\frac{\left(\dfrac{N_i}{N_k}\right)^{S}}{\left(\dfrac{N_i}{N_k}\right)^{O}} -1 \eqno (8)
$$
Again, when $E^{O}_{i,k}<0$, this corresponds to a depletion of the isotope $i$ relative to $k$. 

If the initial ratios in the star are $(N_i/N_k )^O$, then we define 
$$
\left(1+\alpha_{i,k}\right) \equiv \left(N^O_i/N^O_k\right)/\left(N^{\odot}_i/N^{\odot}_k\right) \eqno(9)
$$
As we assume that $n_i/n_k$ in the returning processed parcel is the same for all $s$ processing in an AGB star, the enrichment rule stays the same. Hence, equations from (2) to (5) can be generalized to an arbitrary initial state (O):
$$
E^O_{i,k} = \left(1-f_k\right)\left[\left(n_i/n_k\right)/\left(N^O_{i}/N^O_k\right)-1\right] = \left(1-f_k\right)Q^O_{i,k} \eqno(10)
$$
Further, as $n_i/n_k$ is taken as fixed, then:
$$
Q^O_{i,k}\left(1+\alpha_{i,k}\right) = Q^{\odot}_{i,k}-\alpha_{i,k} \eqno(11)
$$
When measurements are made of stellar debris, the standard reference state for isotopic ratios is solar. Thus all measurements so represented are $(N_i/N_k)/(N_i/N_k)^{\odot}$, independent of the actual initial value in the source. The effective \textit{calculated} enrichment factors are thus:
$$
E^{Calc}_{i,k}=\left(1+\alpha_{i,k}\right)E^{O.mix}_{i,k} + \alpha_{i,k} \eqno(12)
$$
It follows that:
$$
\left(E^{Calc}_{i,k}-\alpha_{i,k}\right)/\left(E^{Calc}_{j,k}-\alpha_{j,k}\right)=\left(Q^{\odot}_{i,k}-\alpha_{i,k}\right)/
\left(Q^{\odot}_{j,k}-\alpha_{j,k}\right) \eqno(13)
$$ 
This is the explicit, general relationship between the calculated enrichments from laboratory measurements to any possible AGB source, assuming the approximations to be adequate. It relates the measured isotopic ratios to a standard reference $s$ process source $Q^{\odot}_{i,k}$, as given above and to the shifts in the initial composition of the source. Note that there are as many values of $\alpha_{i,k}$ as there are isotopic pairs, so the system is not determinate. However, if one of the $\alpha$'s can be estimated, then all of the others may be computed from the calculated delta values for a sample and some standard $Q^{\odot}_{i,k}$, such as given in Table \ref{tab1}.

Comparison of stellar models with results for a sample ($X$) measured in the laboratory requires some special care. The range of effects is enormous and the measured laboratory results are presented in different forms. They are given in units 
of $\delta$ (parts per mil) or $\epsilon$ (parts per 10$^4$) as defined below:
$$
\delta^X_{i,k} \equiv 10^3\left(I_{i,k}^{Sample}/I_{i,k}^{\odot}-1\right)
\equiv 10^{-1} \times \epsilon^X_{i,k} \eqno(14)
$$
so that:
$$
E_{i,k}^{\odot,Calc} = 10^{-3} \delta^X_{i,k} = 10^{-4} \epsilon^X_{i,k} \eqno(15)
$$
Equation (13) can be then rewritten, in terms of \oq delta values\cq, as: 
$$
\left(10^{-3}\delta^{Calc}_{i,k}-\alpha_{i,k}\right)/ \left(10^{-3}\delta^{Calc}_{j,k}-\alpha_{j,k}\right)=\left(Q^{\odot}_{i,k}-\alpha_{i,k}\right)/
\left(Q^{\odot}_{j,k}-\alpha_{j,k}\right) \eqno(16)
$$
Here $I_{i,k}$ is the ratio of the ion beams of isotopes $i$ to $k$ measured in a sample, corrected by a discrimination factor (1+$D_{i,k}$). For ion-probe data, the discrimination factor is typically determined from the measured ion beam ratio $I^{\odot}_{i,k}$ in a terrestrial sample. It is usual, in many cases, that instrumental discrimination factors are determined by one pair of isotopes $(u, v)$ relative to the nominal Solar standard (usually a terrestrial material) and appropriately applied to other isotope pairs assuming some functional dependence on mass. It is important to note that the experimental errors for the delta representation used for Sputtered Ion Mass Spectrometer (SIMS) or nano-scale SIMS (nano-SIMS) are typically given as one sigma. In contrast, the smaller effects measured by Thermal Ion Mass Spectrometry (TIMS) or Ion Coupled Plasma Mass Spectrometry (ICPMS) are two sigma errors. These effects are much smaller, but are more precise. They often represent a normalization to a terrestrial ratio for two isotopes to correct for in-run fractionation. When isotopic ratios are shifted, the normalization can lead to spurious effects.

The problem to be addressed is to relate a set of observations for many elements and isotopes in an individual CIRCON to the composition of a single stellar envelope at some stage \oq $S$\cqb that would produce such a grain \citep[see review by][]{zinner14}. That is, one seeks to find a stellar model where at some stage $E^{\odot,S}_{i,k}$ is related to the observed value in the grain. 


In this case of laboratory measurements of an exotic material, we take $E_{i,k}^{\odot,Calc}$ from equation (15). For the case of processed, but not completely homogenized samples of solar system material, one tries to relate the effective average of some sources $\langle E^S_{i,k}\rangle$ to the observed isotopic composition by some undetermined scaling factor and seek to infer which sources are plausible. Large effects for different elements may be more definitive on a single object. In all cases, the variation in the initial state of stars relative to the Solar reference values must be considered.
 
We will first, for simplicity, calculate the nature and magnitude of effects for Cr, Fe, and Ni and then discuss the other Fe group elements. We will initially assume in our numerical calculations: $N^O_i/N^O_k = N^{\odot}_i/N^{\odot}_k$ (which is not in general true). 

We will show the high degree of enrichment for $^{54}$Cr, $^{58}$Fe, and $^{64}$Ni. For Ti, a deformed $J$-shaped trend is exhibited, with a major effect in $^{50}$Ti.

As the numerical calculations used solar values, we must consider the extent to which variations in the original stellar source would affect the results. 
From Figure \ref{q_values}, we note that the $Q_{i,j}$ values for $^{54}$Cr, $^{58}$Fe, and $^{64}$Ni are large, but are small for all the other Cr, Fe, Ni isotopes. As a result, changes in $Q_{i,k}$ due to changes in the initial composition (say, $\alpha_{i,k} \sim$ 2) will not greatly alter the general isotopic pattern, with the dominant effects being at $^{54}$Cr, $^{58}$Fe, and $^{64}$Ni. Hence, the resulting {\it patterns} for Cr, Fe, and Ni are robust and will be roughly fixed for the AGB models considered here, independent of small shifts in the stellar initial abundances. 

\section{Full isotopic results for AGB stars: Cr, Fe, and Ni products}
In this section we present the detailed results for the isotopic composition of the envelope, represented by $E^{\odot}_{i,k}$, over all pertinent stages of a model star evolution, from the first formation of a $^{13}$C pocket until the last $s$-process event, prior to the transition of the star into a planetary nebula (see Figure \ref{over}), which leaves a white-dwarf remnant. These are the results of the full stellar model calculation and not the partial ones, referring to the He shell, given above in Figure \ref{q_values}. It is readily seen that the patterns for all cases are very close to that obtained for a single $s$-processed parcel (see Figure \ref{q_values}), but that there are small to significant shifts depending on the details of the particular stellar mass and dredge-up episode. We will discuss the effects of different initial states later in section 5. The degree of enhancement increases with the number of pulses due the repeated neutron processing and also to the decrease in the remaining mass of the envelope. 
In Figure \ref{over} we plot the data normalized so that $E_{i,k}$ is set to one for a relevant normalization isotope (e.g. the isotope $^{64}$Ni for nickel), using for reference an isotope that is much enhanced. The figure inset gives the scaling factors for each case. Note that there are a wide range of scaling factors, but that in general, the forms of the curves for all cases are similar. This is a reflection of the nature of $s$-processing in the Fe-peak, as shown in section 2.3.1, and which generates isotopic shifts for Cr, Fe, and Ni with clear major excesses in $^{54}$Cr, $^{58}$Fe, and $^{64}$Ni relative to the index isotopes and only small effects for the other isotopes. The circumstellar condensates with Cr, Fe, and Ni formed around AGB stars will have a wide range of enrichments (as revealed by the scaling factors in Figure \ref{over}) depending on the stage of mass loss and dust formation and on the stellar mass and metallicity.

Note in particular that an enrichment of $^{54}$Cr/$^{52}$Cr is found that can run up to $E_{54,52} \sim 1$ for the bulk envelope at late stages, for the low metallicity case. We notice that, if some material is ejected without full dilution in the envelope, at any stage, then the effects can be very much larger, with the isotopic pattern being fixed by the purely $s$-processed parcels discussed in the previous section. 

\begin{table}[t!!]
\begin{center}
\caption{\label{tab2}Ratios of $E_{i,k}$ for $^{54}$Cr, $^{58}$Fe, $^{64}$Ni, $^{42}$Ca, and $^{29}$Si with respect to $^{50}$Ti. Calculations refer to the last thermal-pulse of 1.5 and $3M_{\odot}$ stars assuming two different metallicities (solar and 1/3$Z_{\odot}$).}
\begin{tabular}{ccccccc}
\hline
\hline
$E_{(i,k)}$ Ratio & Element Ratio & \multicolumn{2}{c}{$1.5M_{\odot}$} & & \multicolumn{2}{c}{$3M_{\odot}$}\\ 
\cline{3-4}\cline{6-7}
 & & $Z_{\odot}$ & $Z_{\odot}/3$ & & $Z_{\odot}$ & $Z_{\odot}/3$ \\
\hline
$E_{(54,52)}/E_{(50,48)}$ & Cr/Ti & 0.35 & 0.16 & & 0.49 & 0.27 \\
$E_{(58,56)}/E_{(50,48)}$ & Fe/Ti & 1.65 & 1.00 & & 3.57 & 1.90 \\
$E_{(64,58)}/E_{(50,48)}$ & Ni/Ti & 7.49 & 2.16 & & 5.33 & 3.08 \\ 
$E_{(42,40)}/E_{(50,48)}$ & Ca/Ti & 0.22 & 0.14 & & 0.22 & 0.18 \\
$E_{(29,28)}/E_{(50,48)}$ & Si/Ti & 0.14	& 0.05 & & 0.09 & 0.03 \\
\hline
\end{tabular}
\end{center}
\end{table}


\section{Meteoritic data and comparison with model predictions}
It is important to note that, while excesses in an isotopic ratio relative to the solar ratio may be explained by addition of that isotope to a Solar mix, negative values require that something is being added having an isotopic abundance pattern for several isotopes which is quite distinct from the Solar pattern. This is a well known fact that is often ignored. For all ion-probe data on CIRCONs where the isotopic shifts are large (many hundreds per mil), we believe that all the data processing that has been applied is to correct for instrumental discrimination effects, by using the isotopic ratios of an isotope $i$ relative to some index isotope $k$ through the determination of that ratio in a terrestrial standard for all isotope pairs. This correction is then applied to the measurements of all data. 
In contrast, the data on CAIs, which are far more precise ($\epsilon$ units), are most often normalized to a pair of isotopes to correct for instrumental fractionation. This procedure effectively requires one pair of isotopes to have the terrestrial ratio and thus affects the pattern in all the other ratios. This difference in the treatment of the experimental data must be kept in mind.

\subsection{Cr}
The discovery of isotopic anomalies in Cr by \citet{ba84} and \citet{bl88} established the presence of small $^{54}$Cr/$^{52}$Cr excesses (of the order of $\sim 6\times 10^{-4}$, or 6 $\epsilon$ units) in CAIs from the Allende meteorite. \citet{pap} then measured Cr in the two \oq fractionation and unknown nuclear anomalies\cqb (the so-called FUN) inclusions, (EK1-4-1 and C-1) that had previously been discovered to have very large isotopic anomalies in many elements. All these mass spectrometric data are normalized to $^{50}$Cr/$^{52}$Cr to match a terrestrial standard. Considering the magnitude of isotopic shifts in CAIs, he found a very large excess in $^{54}$Cr (48 $\epsilon$ units) and a large excess in $^{53}$Cr (16 $\epsilon$ units) in EK1-4-1. However, for C-1 the effect is a large depletion in $^{54}$Cr ($-$24 $\epsilon$ units) and data on non-FUN samples displayed subdued effects that were intermediate in behaviour (non-FUN samples of CAIs are also shown in Figure 1 of that paper).

The issue is then to understand whether any of these patterns can be plausibly explained by an AGB model. In general, most CAIs exhibit excesses in $^{54}$Cr. Analyses of bulk meteorite samples (CV, C and CM types) exhibit no effects at about 1 $\epsilon$ unit level. However, an investigation by \citet{rotaru} showed that these meteorites, if subjected to acid leaching, yielded widely variable Cr isotopic compositions in any given leach sequence on a single meteorite. They found, for the C1 chondrite Orgueil, a range of values from $\sim -9$ to $\sim + 100$ for $\epsilon_{54}$ in different leaches. In most cases, the results showed strong excesses of $^{54}$Cr. \citet{podosek} sought to find the carrier grains of anomalous Cr by a stepwise dissolution process and found $^{54}$Cr excesses of up to $\sim$ 210 $\epsilon$ units, but no identification of carrier grains. It is clear that there are materials of highly divergent nuclear sources in any bulk sample. Several groups have carried out searches for the carrier grains of the anomalous Cr. \citet{dauphas} presented an extensive study searching disaggregated and treated material for carrier grains from 1,400 to 30 nm in size. There were clear excess in $^{54}$Cr, but not of very large amounts (up to 17 $\epsilon$ units). Only one exceptional grain with $E_{54,52} = 2.9 \pm 0.33$ was found, which was attributed to a supernova source on account of its very large excess. In a study by \citet{qin} on insoluble residues with Cr-rich grains ($\sim 200$ nm), they found three grains that are enriched up to 1.5 times the solar value, and concluded that this must be the result of supernova debris injected into the protosolar cloud. These grains are plausibly circumstellar condensates; however, there are no measurements made on identified circumstellar grains that are clearly related to AGB sources; the measurements themselves are very difficult to perform and have large errors. The Cr data obviously show clear effects, with frequent $^{54}$Cr excesses and some $^{54}$Cr deficits; but full measurements for all Cr isotopes are limited and no data are available on well-defined circumstellar condensates that are clearly related to AGB stars.

\subsection{Fe}
Extensive data on a variety of bulk meteorites and some Fe-Mg chondrules from Allende show no shifts in the Fe isotopes, including $^{58}$Fe. The only available data on Fe anomalies is by \citet{volk}. The results, obtained on the same solutions of EK1-141 used for Cr measurements \citep{pap} are of high precision and normalized to the solar value of $^{54}$Fe/$^{56}$Fe. They show clear, large excesses of $^{58}$Fe of 292 $\epsilon$ units and no shifts for $^{57}$Fe. The C-1 sample shows instead only marginal enrichments in $^{58}$Fe; other CAIs exhibit small, but clear excesses in $^{58}$Fe. The data on Fe in CIRCONs are limited. The first reports were by \citet{hop}. Using an IMS-3F, their focus was on $X$-grains that are associated with SNe. \citet{hop} only measured $^{54}$Fe/$^{56}$Fe and did not detect any effects. These are difficult measurements and they noted limits due to several isobaric interferences including $^{54}$Cr. A recent report by \citet{marhas} presented extensive data on Fe and Ni isotopes in SiC grains using a Cameca NanoSims. Again, there are limits on which isotopes could be measured due to isobaric interferences, in particular for $^{54}$Cr, $^{58}$Fe, and $^{64}$Ni. As a result only $^{54}$Fe/$^{56}$Fe, $^{57}$Fe/$^{56}$Fe, $^{60}$Ni/$^{58}$Ni, $^{61}$Ni/$^{58}$Ni, and $^{62}$Ni/$^{58}$Ni were measured with corrections for $^{54}$Cr to $^{54}$Fe and for $^{58}$Fe to $^{58}$Ni, assuming terrestrial values for the corrections. The focus of this work was again on $X$-grains associated with SNe. Large excesses were found for $^{57}$Fe and $^{61}$Ni, $^{62}$Ni in some $X$ grains. Extensive measurements were also made on Mainstream grains associated with AGB stars. These showed no large shifts (see their Figures 4 and 5). The major emphasis in \citet{marhas} was to identify the effects they measured with selected zones in a reference SNeII model. This approach assumes that different zones are sampled to produce the isotopic pattern and impart it in to a SiC grain. These workers also presented the effects of $s$-processing Fe and Ni in the envelopes of AGB stars. However, they present no calculations for $^{54}$Cr, $^{58}$Fe or $^{64}$Ni. We presume that these were ignored because they could not be measured. As we showed in section 2, these are the isotopes that will exhibit, by far, the largest effects from $s$-processing in AGB stars; as a consequence, the results on main stream grains in the \citet{marhas} report do not either support or negate the findings presented in the present work.

\subsection{Ni}
The discovery of small (a few $\epsilon$ units) anomalies in Ni was found by \citet{bl88}. The largest effect was on $^{64}$Ni (about 3 $\epsilon$ units). Measurements of bulk meteorites \citep{regelous} showed no isotopic shifts at a high level of precision ($< 0.1$ $\epsilon$ units) for many meteorites, but no data on $^{64}$Ni were present. 

An important, critical measurement on Ni was made by \citet{cpap}. These workers showed endemic $^{64}$Ni effects in CAIs which ranged from $+55$ $\epsilon$ units (Egg-3) to very low levels ($\sim 1$ $\epsilon$ units) in some other samples. The data were {\it normalized} to the $^{58}$Ni/$^{61}$Ni value for terrestrial Ni. The effects on all other isotopes were small. One sample showed a distinctly negative value ($\sim 1$ $\epsilon$) at $^{60}$Ni (with their normalization). The observation on Egg-3 was the focus of our interest as it showed unambiguous nuclear effects with a clear pattern for this key \oq Fe group\cqb element. Again, there are no Ni isotopic data on CIRCONs, other than the results reported by \citet{marhas}, which, as indicated above, are not adequate for the problem under discussion.

\subsection{Ti}
There are abundant isotopic data on Ti, for solar-system processed samples (CAIs and many bulk meteorites). The patterns are complex and varied. Many publications on Ti isotopes in processed solar system material are available, but the results we mention below are typical of the whole range observed. It is often found that there are well defined excesses at $^{49}$Ti, $^{47}$Ti, and $^{50}$Ti, with the $^{50}$Ti anomalies being larger than for $^{49}$Ti (as found in EK-1-4-1). There are also samples with distinct, large deficits in $^{50}$Ti and smaller, but clear, deficits in $^{49}$Ti (C-1): \citep[see][with a normalization to solar $^{48}$T1/$^{46}$Ti]{niederer1}. Absolute isotopic ratios were obtained using a modified technique that eliminates the normalization for instrumental fractionation \citep{niederer2}. These workers showed that at least three, and possibly four stellar sources must have contributed to the five Ti isotopes to make up the samples of processed materials. An extensive discussion of possible stellar sites and sources was also presented by them in trying to identify the compositions of the exotic components. It is evident that the processed solar materials do not yield a direct tie in with individual possible stellar sources.

Because of the problem of complex mixed sources, we turn to the Ti data in CIRCONs, with the hope of finding some regularity using known nucleosynthetic signatures of AGB stars. The AGB processing for a single grain then depends on the stellar evolution and the initial composition of the parent star. In particular, the variations observed in Ti isotopes found in oxygen-rich CIRCONs \citep{choi} suggested that the Ti isotopic compositions follows the pattern generated by AGB processing, following on the calculations by Gallino and co-workers. \citet{nittler08} presented an extensive study of Al, Ca, and Ti in CIRCONs. They showed very strong evidence that most of these grains were formed in red giant and AGB parent stars. They reported Ti data on three oxide grains that showed small isotopic effects, but with divergent patterns. These grains must have been produced in a regime where C/O$<$1, implying a low mass star prior to C enhancement.

Extensive studies of the isotopic composition of many elements in SiC circumstellar condensates \citep{zinner14} were presented by several workers; see e.g.  \citet{amari1}; in particular, see their Figure 5. See also \citet{amari2}, in particular see their Figure 4. The data exhibit a wide and diverse array of very different patterns; these last have often been described as being \oq V\cq-shaped, but with large variations, sometimes negative. We will rather use a \oq J\cqb shape as a descriptor, because that is the pattern which would be produced in AGBs from an initially Solar mix (see Figure \ref{ti-1}), in seeking some hint as to which examples from the zoo of Ti isotopic patterns might serve as a guide. We note that several samples show \oq J\cqb shape patterns in \citet{amari1}; the same is true for \citet{huss}, see in particular their Figures 2, 3, and 11. The patterns of many SiC grains are quantitatively indistinguishable from those of AGB models for an initial solar isotopic abundance. We show the measurements by \citet{amari1} for KJGM1-158-5, which has a very large enrichment in $^{50}$Ti, in Figure \ref{ti-1}. For comparison, we include the results for $Q_{i,48}$ solar, normalized to the measured $\delta_{50,48}$ value. It can be seen that within the errors, the results are indistinguishable,  if one considers the errors in both $^{50}$Ti/$^{48}$Ti and $^{49}$Ti/$^{48}$Ti.  

We now consider a wider number of samples, some of which have only more modest effects in $\delta_{50,48}$ (down to 200 per mil). Using data from \citet{amari1,amari2} and grains 101, 58b, and 110 from \citet{huss}, we show in Figure \ref{ti-2} the results, normalized to the measured $\delta_{50,48}$, including also error bars. Again a set of normalized reference curves using the $Q_{i,48}$ values are shown. Some samples, such as KJGM4C-314-3, appear to be widely deviant with a $^{49}$Ti shift above that of $^{50}$Ti. However, if one considers the one-sigma error bars of 20\% for $\delta_{49,46}$ and of 24\% for $\delta_{50,46}$, there is no clear discrepancy. From these observations we conclude that a large fraction of these grains show clear evidence of $s$-processing. However, there are many cases that also show clear disagreement with significant $s$-processing. These differences, shown in Figure \ref{ti-2}, would correspond to shifts in initial isotopic abundances of 100 per mil for some isotopes which are not enriched in the $s$-process calculations presented here.

We are then left with the issue of what the variations are due to. Many of the samples showing clear deviations from AGB models still exhibit aspects of the \oq J\cqb or \oq check\cqb pattern. Note that for $^{46}$Ti (which is 8\% of the total Ti), a 5\% increase in the $^{46}$Ti/$^{48}$Ti ratio would give (relative to Solar) a 5\% shift in $^{46}$Ti/$^{48}$Ti and change the \oq J\cqb pattern produced by $s$-processing into a \oq V\cqb pattern. There is a very strong indication that many of the variations observed represent $s$-processing, but from somewhat different initial isotopic compositions. When the $s$-process effects are large (due to grain formation in very advanced evolutionary stages), the Ti isotopic pattern closely approaches that of pure $s$-processing. When instead grains form at earlier dredge-up episodes, then the initial composition dominates.


\subsection{Si}
We now consider Si isotopic shifts in CIRCONs in the light of the above discussion of Ti which shows clear evidence of $s$-processing, but with some intrinsic variability in the initial values due to granularity in the source ISM. The data on SiC grains (main stream and A, B grains) associated with AGB stars exhibit a clear trend in the isotopic ratios of Si, as found by other workers \citep[see Figure 3 in][]{zinner14}. The fundamental issues on the isotopic composition of Si in the ISM were fully recognized and discussed by \citet{TIM96}. These workers discussed the diversity of SNe sources and the possibility that these effects could be used to trace earlier galactic evolution. The observed trend expressed in terms of $\delta^{\odot}_{30,28}$ versus $\delta^{\odot}_{29,28}$ has a slope close to (or larger than) unity, with some points displaced in the direction of a line of much smaller slope ($\sim 0.3$). The slope of $\sim$ 0.3 is calculated for a model with $1/v$ dependence for neutron capture cross sections. This inconsistency of predictions with the main trend of high slope was recognized by many authors. Some significant efforts have been made to explain these results. We note that the cross sections do not follow the $1/v$ rule and are subject to large uncertainties, especially for $^{30}$Si (Franz K\"appeler \& Iris Dillman, private communication). Hence the slope predicted for AGB stars may be more complex than previously assumed. It is dependent on the effectiveness of \ctb burning and on the extension of the \ctb reservoir. For small pockets, the slope is small ($\lesssim 0.5$) and determined mainly by neutrons from the $^{22}$Ne($\alpha$,n)$^{25}$Mg reaction. For large extensions of the pocket the slope increases substantially, up to about 0.8 in our models. In no case can large enrichment factors for $^{29}$Si and $^{30}$Si be produced in AGB envelopes, as the $^{28}$Si neutron capture cross section is small (this is also the case for Ca isotopes). While more advanced AGB computations may change the results, we consider here that the SiC measurements are plausibly explained by other important factors. In particular, the possible intrinsic local variability in Si in the local ISM from which the grains formed must be considered. It is of importance to note that the relative enrichment of Si and Ca effects from AGB processing as compared with Ti, here is rather low (see Table \ref{tab2}). In contrast, $^{58}$Fe and $^{64}$Ni should exhibit larger effects than $^{50}$Ti. For $^{54}$Cr there should be smaller effects than for $^{50}$Ti. Thus, in consideration of the small variability (10\% to 20\%) in the ISM isotopic composition estimated from the Ti data, we expect that the coupling of Si data with Ti due to AGB processing will be very weak. The dominant effect for Si is due to intrinsic variability in the contributions from distinctive SNe sources. Coupling the results for both Si and Ti caused many workers to invoke a model of Galactic Chemical  Evolution to interpret the observations.
The original proposal to explain the O isotopic data in CIRCONs by Galactic Chemical Evolution was made by \citet{nit1} and discussed further by \citet{nittler1} in a stochastic model, considering mixing from different sources.

Efforts to relate the Ti and Si effects to Galactic Chemical Evolution resulted in some difficulties. As pointed out by \citet{amari1} and by \citet{nittler1}, the correlation lines for Si and Ti isotopes for the Galactic Evolution trend do not pass through the solar value. Further, \citet{nittler1} pointed out that the Si trend could be explained if the role of SNeIa ejecta in producing $^{28}$Si was larger than assumed in the previous years. In an important paper by \citet{lugaro}, after the work of \citet{TIM96}, the following points were made: i) that there was local heterogeneity in the ISM at the time when the \oq main stream\cqb grains were formed; and ii) that there are shifts to be expected in the Si isotopic abundances in the ISM due to intrinsic variability in the ratios produced by core collapse SNe and by SNeIa. These workers still argued that there are effects of Galactic Chemical Evolution, but considered the possibility that there is intrinsic heterogeneity in the ISM at essentially all times.

The wide possible range in $^{29}$Si/$^{28}$Si and $^{30}$Si/$^{28}$Si presented by \citet{lugaro} was based on the detailed SNe yields by \citet{ww} and is seen in a subsequent report by \citet{rausch}. We note that recently published yields of the Si isotopes for SNeIa clearly indicate that, indeed, $^{29}$Si and $^{30}$Si are grossly under produced as compared to $^{28}$Si in the ejecta \citep{bp12}. SNeIa are major source of Fe. The above authors showed that $^{29}$Si/$^{28}$Si $\sim$ 1.96$\times$10$^{-3}$ and $^{30}$Si/$^{28}$Si $\sim$ 3.6$\times$10$^{-3}$. These are about 1/10th the solar values. The slope of a line connecting Solar and these SNeIa values is 0.93 in the $\delta^{\odot}_{30,28}$ versus $\delta^{\odot}_{29,28}$. 

About 15$-$45\% of the $^{28}$Si in the bulk Solar inventory is from SNeIa sources \citep{lugaro} and all of the remainder is from SNeII. The various SNeII sources (depending on mass) show a wide spread in Si isotopic abundances \citep{ww, rausch}. Calculating a grand canonical average using an initial mass function that gives the large scale Solar Si isotopic ratios provides a plausible base line as a guide. However, assuming Solar abundances as a universal rule is not plausible. It is evident that the bulk Solar value is the result of a blend of diverse sources with much of the $^{29}$Si and $^{30}$Si coming from very massive stars \citep[$>$ 30\ms, see][]{cl13}. If we consider inhomogeneities in the ISM resulting from rather small variations in the mixing ratios, then a 1-10\% change in the contribution of SNeIa sources to a local star formation region would result in a shift of 100 per mil for Si isotopes. This is essentially the trend that is observed for the mainstream and the A and B grains \citep[see][and references therein]{amari1}. However, one must further note that the Si data show that the $Y$ grains and the $Z$ grains are displaced to the right of the main array (which has a slope of $\simeq$ 1). It is possible that this displacement is due to AGB processing of materials that were on the slope $\sim 1$ line. In general, the $X$ grains (which have very negative deltas) are recognized as having rather direct connections with SNe condensates. 

The aforementioned study by \citet{nittler1} of a stochastic sampling of SNe
ejecta from different sources found that this approach does not reproduce the strong correlation between $^{29}$Si/$^{28}$Si and $^{46}$Ti/$^{48}$Ti ratios observed in CIRCON grains. However, this approach attributes both Si and Ti isotopic effects to SNe. If the dominant effects for Ti in mainstream  CIRCONs are due to AGB re-processing and the Si effects are from SNe, then this correlation is not applicable. 
The O isotopic patterns are in many cases not well understood and may indeed require SNe contributions.  More work is needed on this subject with full attention given to intrinsic heterogeneities in the ISM.

We infer that the Si isotopic data of CIRCONs found in meteorites are plausibly explained by modest variability in the isotopic composition of Si in star forming regions.

\subsection{Ca}
Measurements of all the Ca isotopes are quite difficult (particularly for the rare isotope $^{46}$Ca, which is only about $3\times 10^{-2}$ of the total Ca). There are also isobaric interferences from $^{46}$Ti and $^{48}$Ti for the corresponding mass numbers. Non-linear isotopic anomalies of Ca (not attributable to mass dependent isotopic fractionation) in CAIs were first discovered by \citet{lee1}. Normalizing their data to solar $^{40}$Ca/$^{44}$Ca, they found that $^{48}$Ca was enriched in EK141 by $+ 13.6$ per mil. For sample C-1 they found a deficit of $\sim -3.0$ per mil in $^{48}$Ca. No effects were detectable in $^{46}$Ca. \citet{np84} determined absolute isotopic ratios for CAIs, eliminating the normalization procedure to correct for instrumental fractionation effects. They confirmed the effects reported for EK141 and C1; they also found a CAI with very large excesses in $^{46}$Ca ($\sim 7$ per mil) and with no effects in $^{42,43,48}$Ca. \citet{papbr}, in a study of Mg, Ti, Cr, and Ca in CAIs, found clear, very large deficits in $^{48}$Ca ($\sim 31$ per mil) and no detectable effects in $^{46}$Ca. These samples also showed large correlated deficits in $^{49}$Ti, $^{50}$Ti, and $^{53}$Cr (see their Figure 1). These reports clearly demonstrate significant isotopic effects in Ca in processed solar materials and represent the largest effects so far found in them. A recent work of very high precision was carried out by \citet{huang} on CAIs and found that there were no detectable shifts in $^{43}$Ca/$^{44}$Ca or $^{40}$Ca/$^{44}$Ca (below $\sim 1$ $\epsilon$ units), but several samples showed clear excesses of $^{48}$Ca ($\sim$ several $\epsilon$ units). In their study, these authors used a double spike technique so that some of the data are not normalized to correct for instrumental fractionation. The absolute $^{44}$Ca/$^{40}$Ca and $^{44}$Ca/$^{42}$Ca data are very precise ($\pm 0.06$ per mil). However, the $^{48}$Ca/$^{44}$Ca ratios were calculated assuming the terrestrial value. This means that even the small errors in $^{44}$Ca/$^{40}$Ca and $^{44}$Ca/$^{42}$Ca would propagate into shifts of over 1.2 $\epsilon$ units in the $^{48}$Ca/$^{44}$Ca ratio, which is not much less than the effects reported considering the listed errors. No measurements were made of $^{46}$Ca for the reasons given above. They also found clear evidence of mass-dependent fractionation in both CAIs and bulk meteorite samples \citep[see also][]{huang2}. \citet{moynier} leached samples of a C-1 chondrite searching for carrier grains of $^{54}$Cr and also measured Ca. They found widespread small ($\sim 1$ $\epsilon$ units) variations in $^{40}$Ca and examples of $^{46}$Ca deficits in one case and a couple of cases where there was clear small $^{48}$Ca excess ($\sim ~ 3$ $\epsilon$ units). It is important to note that large errors are often found for $^{46}$Ca measurements. They further noted that $^{54}$Cr effects were not correlated with those on $^{48}$Ca. \citet{chen} carried out an extensive study of $^{48}$Ca anomalies in a wide variety of bulk differentiated meteorites, again using a normalization to $^{42}$Ca/$^{44}$Ca following \citet{russell}. They confirmed the ubiquitous, but small, isotopic heterogeneity in planetary bodies of the Solar system for $^{48}$Ca and also reported one sample with $^{46}$Ca excess far outside {\bf their typical} experimental error. In their discussion, they focused on SNe sources for the $^{48}$Ca effects, since $^{48}$Ca is produced in SNe. 

The isotopic data on Ca for CIRCONs are quite limited. There are no measurements of $^{46}$Ca and very few for $^{48}$Ca. Note that sputtering ion mass spectrometry, which does not use preparatory chemistry to remove interfering elements, requires that interferences must be resolvable and identifiable. The interferences from $^{46}$Ti and $^{48}$Ti make such measurements very difficult for a rare isotope like $^{46}$Ca.

\citet{choi} measured $^{42}$Ca/$^{40}$Ca and $^{44}$Ca/$^{40}$Ca in two CIRCON oxide grains and found them to be the same as the solar value, but with large uncertainties. \citet{choi1} subsequently measured two more CIRCON oxide grains and found $^{42}$Ca/$^{40}$Ca, $^{43}$Ca/$^{40}$Ca, and $^{44}$Ca/$^{40}$Ca to be normal within errors. \citet{nittler08} presented a very extensive and thoughtful study of Al, Ca, and Ti in a large number of CIRCON oxide grains associated with AGB sources. They presented data on $^{42}$Ca/$^{40}$Ca, $^{43}$Ca/$^{40}$Ca, and $^{44}$Ca/$^{40}$Ca. In five cases they found shifts in $^{42}$Ca/$^{40}$Ca from $+ 249$ to $- 38$ per mil relative to solar. For $^{43}$Ca/$^{40}$Ca they found shifts of $+118$ per mil in one sample and for $^{44}$Ca/$^{40}$Ca positive shifts in some samples going up to $+181$ (see their Figure 5). They fully recognized the possible role of AGB processing and showed the pattern to be expected for a 2 \msb star. Our calculations of both pattern and yield are in accord with their report. They did not include $^{46}$Ca and $^{48}$Ca, but they recognized that $^{46}$Ca would be produced. They further discussed the relationship between $^{42}$Ca/$^{40}$Ca and $^{43}$Ca/$^{40}$Ca to be expected from AGB sources. 

In considering the problem of $^{48}$Ca we also recognize that any inventory of this isotope must be due to some SNe sources. However, the existence of both positive and negative effects for $^{48}$Ca (and many other isotopes) proves that any simple addition mechanism involving one isotope is not possible. Inspection of the curves of $Q_{i,40}$ (Figure \ref{q_values}) or of $E_{i,40}$ (Figure \ref{over}) shows that no $^{48}$Ca is produced in AGB processing and that this nucleus is also not efficiently destroyed. We further note that the production of $^{42}$Ca, $^{43}$Ca, and $^{44}$Ca is subdued, but is a natural result of AGB processing. For the case of a Solar initial state in a star, the results enhance $^{42}$Ca relative to $^{44}$Ca. This can be clearly seen in Figure \ref{calcium}, where it is shown that enrichment for $^{42}$Ca is always greater than for $^{44}$Ca. It follows that shifts in $^{42}$Ca/$^{44}$Ca are to be expected. As all of the estimates of $^{48}$Ca excesses/deficits from measurements use a normalization to Solar $^{42}$Ca/$^{44}$Ca, it is clear that any enhancement of $^{42}$Ca/$^{44}$Ca by neutron captures will result in a calculated excess of $^{48}$Ca. Further, if there are small variations $^{42}$Ca/$^{44}$Ca in the ISM at the formation of a star, this will in general result in either excesses or deficits in the calculated $^{48}$Ca/$^{44}$Ca ratio.

From the observational data, we note that all Ca isotopic effects are small, being typically a few $\epsilon$ units in bulk meteorites and most CAIs. In oxide CIRCONs, the effects are less than 100 per mil in most samples. As discussed above, several of the cases show a rather strong resemblance to the AGB model \citep[see also][]{nittler1} and may be explained by variability in the initial inventories of the AGB stellar sources as argued for Ti. 

In considering the Ca problem we focus on three issues:

1) There is no production and no effective destruction of $^{48}$Ca in AGB stars;


2) There is a high degree of enrichment in $^{46}$Ca as a result of AGB processing and very few measurements of this rare isotope. The only clear case is the data on EK141  with only modest excesses of $^{46}$Ca (3 per mil) with a much larger excess in $^{48}$Ca (13 per mil);

3)The effect of $s$-processing on Ca is to produce modest enrichments (see Figures \ref{q_values} and \ref{over}) in $^{42}$Ca, $^{43}$Ca, and $^{44}$Ca with the net effect in all cases to have larger enrichments in $^{42}$Ca relative to $^{44}$Ca.

If the Ca calculation is correct, then there should be very large enrichments in $^{46}$Ca. The large effects we predict for $^{46}$Ca are also to be found in the FRUITY data base. This isotope has only rarely been measured, but no positive evidence of large enrichments in $^{46}$Ca are even hinted at. If the lifetime of $^{45}$Ca is not the laboratory value, with $t_{1/2} =$ 163 days, but is controlled by short lived isomeric states, or by other processes affecting weak interactions in stars \citep{sim}, then the calculated large effects on $^{46}$Ca and the inferred large shifts are not valid. If this is not the case, then we are obliged to infer that there are very large $^{46}$Ca effects to be found or that the $s$-processing fails to explain the Ca data, which would  place in jeopardy the approach presented here of processing of Fe group elements in AGB stars as the cause of observed shifts in these elements. The behaviour of $^{48}$Ca is a matter of some issue. The measurements of $^{48}$Ca in almost all cases depend on a normalization procedure using $^{42}$Ca and $^{44}$Ca. In so far as the material being analyzed has small variability in the Ca isotopes, then the normalization $^{42}$Ca/$^{44}$Ca must remain in doubt. If a sample has been subject to processing in AGB stars, then it will be enhanced in $^{42}$Ca relative to $^{44}$Ca. A 2 $\epsilon$ units enhancement in the $^{42}$Ca/$^{44}$Ca ratio in such material would give a calculated 4 $\epsilon$ units excess in $^{48}$Ca/$^{44}$Ca. We consider that this may be an explanation of some of the reported $^{48}$Ca excesses. However, as pointed out by D.A. Papanastassiou (private communication) if the shifts in $^{42}$Ca/$^{44}$Ca are due to $s$-processing, then there will be distinct effects on $^{43}$Ca that are not found. Some variability in the $^{48}$Ca inventory in the ISM is also a most plausible scenario; however, the issue of the $^{48}$Ca anomaly being due to a normalization calculation procedure requires serious attention. 

We find the disparity between the patterns observed for Ca and the one to be expected from AGB processing of an assumed initial solar composition to be similar to what was cited above for Ti. As the shifts in $^{42}$Ca , $^{43}$Ca, and $^{44}$Ca are small, due to the low neutron capture cross section of $^{40}$Ca, small variations in the initial stellar composition will have much enhanced effects on the patterns observed. The Ca problem was one of the most serious difficulties we had in addressing the role of AGB re-processing. 

\subsection{Zn}
The only data on Zn are for CAIs, reported by \citet{volk1}. Normalizing the data to $^{64}$Zn/$^{68}$Zn, they found an excess in $^{66}$Zn of 16 $\epsilon$ units in EK141 and deficits of $\sim 10 - 20 \epsilon$ units for $^{70}$Zn, with large errors. $^{67}$Zn was not measured for technical reasons. All other CAIs showed normal Zn. They discussed the extent to which these anomalies could be produced in neutron rich NSE synthesis. As $^{68}$Zn is the highest enriched Zn isotope from $s$-processing (see Figures \ref{q_values} and \ref{over}) and all the Zn isotopes are significantly enriched by $s$-processing, any normalization procedure is complex. In the absence of data for $^{67}$Zn in EK-141, a relationship to the $s$-processing model is made difficult. No clear explanation for these results are yet in hand.

\section{Summary and conclusions}
This work was initially directed to establish whether the excesses observed in $^{54}$Cr, $^{58}$Fe, and $^{64}$Ni in some CAIs could be the result of AGB processing of matter from the ISM. Detailed calculations showed that the data could be the result of neutron captures in such stars and further, that the effects on these isotopes were much larger than on any of the other isotopes of these elements. As a result, the large excesses in $^{54}$Cr, $^{58}$Fe, and $^{64}$Ni would be present even if there were substantial variations in the initial isotopic composition of the contributing AGB stars. These \oq Fe group\cqb elements were originally made in SNe, but their products have been subsequently reprocessed in AGB stars, resulting in the observed effects. Assuming this to be generally applicable to other 
\oq Fe-group\cqb elements, we then sought to establish whether this might explain the observations on Ti, Si, and Ca. We therefore expanded the presentation to include the wealth of data on pre-solar circumstellar grains.
 
For Ti, the situation is roughly similar to that for Cr, Fe, and Ni, where the predominant enrichment is in one isotope. However, in addition to large effects predicted for $^{50}$Ti, there is also significant to major production of $^{46,47,49}$Ti relative to $^{48}$Ti. The large effects of $^{50}$Ti produced were used as a guide. Using the important and extensive results on circumstellar condensates (CIRCONs), known from other workers to be associated with AGB sources, we sought to relate the Ti isotopic data to AGB processing. It was shown that a substantial part of the Ti data on CIRCONs show clear effects of $s$-processing in stars having close to Solar initial Ti isotopic ratios. However, as is well known, many variations in the Ti data did not fit into the AGB enrichment pattern, particularly for grains that were not greatly enhanced in $^{50}$Ti. Previous workers had inferred that the scatter in Ti isotopic abundance patterns in CIRCONs must be due to variations in the initial states of the different stellar sources. We are in accord with this, but are of the view that, in general, the Ti data are the result of very substantial AGB processing, superimposed on a base of a variable initial isotopic inventory from the local ISM from which the star formed. The required variability is in all cases small: $\sim$ 10\% in isotopes that are at most $\sim$ 10\% of the total Ti inventory (for example, $^{46}$Ti changing from an 8\% to an 8.8\% of total Ti in the local ISM). The negative effects observed can be explained by an increase in $^{48}$Ti relative to the other isotopes by less than 1\%. 

With regard to the widely recognized $^{48}$Ca anomalies, commonly attributed to SNe, we consider that the assumption of Solar $^{42}$Ca/$^{44}$Ca ratio in normalizing the data could generate apparent $^{48}$Ca /$^{44}$Ca excesses/deficiencies. An AGB model predicts large excesses of $^{46}$Ca which are not observed.
A possible explanation may result from the effects of the high stellar electron densities on weak interactions  \citep{sim}, or from a decay rate of $^{45}$Ca governed by some short lived isomeric state in stellar conditions. Work to clarify these possibilities is in progress.

For Si from AGB sources, the regular trends of $\delta_{30,28}$ versus $\delta_{29,28}$ with a slope of $\sim 1$ found for the \oq main stream data\cq, do not appear to reflect the effects of neutron captures with the standard cross sections or neutron exposures. As noted above, the Si isotopic shifts to be expected from AGB processing are very small compared to those for Ti. Any heterogeneity in the ISM such as found for Ti would overwhelm any Ca shifts, so these two data sets should be considered decoupled. The observed Si-Ti isotopic correlation has earlier been considered by many authors to be the result of Galactic Chemical Evolution. However, considering the heterogeneity in initial states clearly shown by the Ti results and the wide range of Si isotopic abundances from different SNe sources (particularly SNeIa) and the small Si effects from AGBs, we infer that the slope 
$\sim 1$ is simply due to a somewhat variable ($\sim$ 10 to 20\%) contribution of SNeIa to the various local regions where these stars formed. Considering the fact that from 1/2 to 2/3 of all the Fe nuclei must come from these sources, which produce $^{28}$Si and very little $^{29}$Si and $^{30}$Si, this appears to be a plausible cause of the observed trend, as early suggested by \citet{lugaro}. We consider the granularity of the ISM composition to be the greater governing agent for most of the data reported on Si and Ca and any coupling between $s$-processing for Ti, Cr, Fe, and Ni with Si and Ca will be weak, because of the low effects resulting from $s$-processing on silicon and calcium themselves. 

We note that the calculations, both in the stellar models and in the data representations, use Solar reference isotopic values. As derived from equations (9), (12), and (13) the enrichment factor relative to the solar reference (expressed by the factor $\alpha_{i,k}$) is (see equation (16):
$$
\left(10^{-3}\delta^{Calc}_{i,k}-\alpha_{i,k}\right)/ \left(10^{-3}\delta^{Calc}_{j,k}-\alpha_{j,k}\right)= \left(Q^{\odot}_{i,k}-\alpha_{i,k}\right)/
\left(Q^{\odot}_{j,k}-\alpha_{j,k}\right) 
$$
This is calculated for a sample formed in the envelope of a star that did not have an initially solar isotopic composition (see section 2.3.3 for details). This simple relation connects the calculated laboratory data relative to Solar standards to the value that would be in a model star of solar initial isotopic admixture, but is actually offset by the actual initial composition through the abundance factor $\alpha_{i,k}$. This permits of determining the initial composition of the source if one $\alpha_{i,k}$ can be estimated. A shift due to $\alpha_{i,k}$ will be very significant (e.g. for $\alpha_{i,k} = 10^{-2}$, this is a 1\%, or 10$^2$ $\epsilon$ unit shift in the isotopic ratio).

Support for possible isotopic variability may be found in the known variability in chemical abundances for stars with [Fe/H] $\simeq 0$ \citep[see e.g.][]{mai12,martig}. Certainly, 10\% to 20\% shifts must be expected in the ISM. The case for galactic changes will be more clearly defined when the observational data and theory require metallicities much below solar and hence significantly earlier epochs. This may be applicable for some of the oxygen results in the report by \citet{nit1}. The role of heterogeneity that we invoke has been widely known, but only in connection with much larger scale issues of Galactic Chemical Evolution. As noted by \citet{nittler08} (section 4.2.3): \oq The fact that the total spread in the grain data is larger than is expected for nucleosynthesis in a single O-rich AGB star suggests that much of the range indeed reflects variations in initial compositions due to GCE. This is analogous to the case for Si and Ti isotopes in presolar SiC grains \citep{ALE99,ZIN07}\cq. We concur that the variations reveal variable initial compositions; however we believe that they  reflect the intrinsic granularity and heterogeneity of the local ISM from which stars form and, to a lesser extent, some effects of broader-scale Galactic Chemical Evolution. For this last, the issue is to identify CIRCON grains which {\it require} low metallicity to explain their isotopic data.

The ISM does not have an exactly solar isotopic composition, but is heterogeneous both in time and space. 
That there is a difference in the initial composition of the sources has long been recognized by many workers \citep{amari1, amari2, huss}. We agree with previous researchers that the silicon and titanium isotope arrays exhibited by presolar SiC grains reflect primarily the initial compositions of the parent stars, overprinted by a component of $s$-process nucleosynthesis \citep[see][]{amari1,huss}. However the small $s$-components are the effect of early stages of dredge up, which emphasize the inherited component; while more advanced stages are dominated by the $s$-process itself.

Measurements of the isotopes $^{54}$Cr, $^{58}$Fe, and $^{64}$Ni are very difficult using SIMs techniques, because of the isobaric interferences. However, some aspects of the AGB effects can be reconstructed with complete data sets of Ti, Cr, Fe, and Ni on single samples of CAIs. Insofar as the isotopic effects are sufficiently large, it should be possible to establish whether the predicted isotopic correlations are really found. For example, the absence of $^{58}$Fe in the presence of excess $^{64}$Ni or of high $^{50}$Ti values would cast serious doubts on the approach presented here.

A careful assessment should now be made of the codes for $s$-process 
in AGB stars with a focus on the Fe group elements, so that errors in the 
nuclear physics or the codes are not propagated in our models. More broadly, 
the general consequences for all nuclides of the results of re-processing 
ISM matter in AGB stars requires careful attention for all affected nuclei,
including radioactive isotopes. Attempts at identifying contributions from Galactic
Chemical Evolution (not from local granularity) should be focused on CIRCONs
grains that may reasonably be considered to come from stars with [Fe/H] values
well below zero. Very high enrichments of $^{50}$Ti may aid in identifying such 
grains. Further work in this field may also be directed to look for correlations 
between some of the signatures discussed in the present paper and the several 
recent reports on very small isotopic variations in many heavier elements (eg Mo, 
Ru, Hf, and W), which show $s$-process excesses or deficits in leaches of various 
meteoritic materials. The problems associated with leaching experiments for 
different chemical elements will of course remain.

{\bf Acknowledgments.}
G.J.W. would like to acknowledge the Epsilon Foundation for continuing support and NASA (grant NNX12AJ01G to A. Krot, University of Hawaii) for partial contribution. This research was co-funded by the Italian INFN, \textit{Istituto Nazionale di Fisica Nucleare}, through the ASFIN collaboration. Thanks are due to INFN also for financial support to O.T.. The authors are greatly indebted to the knowledgeable and unknown reviewer who made very valuable and constructive comments which led to a much improved manuscript. These results were first reported at the R.J. O'Connell Symposium at Harvard, September 2014.


\newpage
{\Large{\bf Figure Captions}}

\begin{figure*}
\includegraphics[width=1\textwidth]{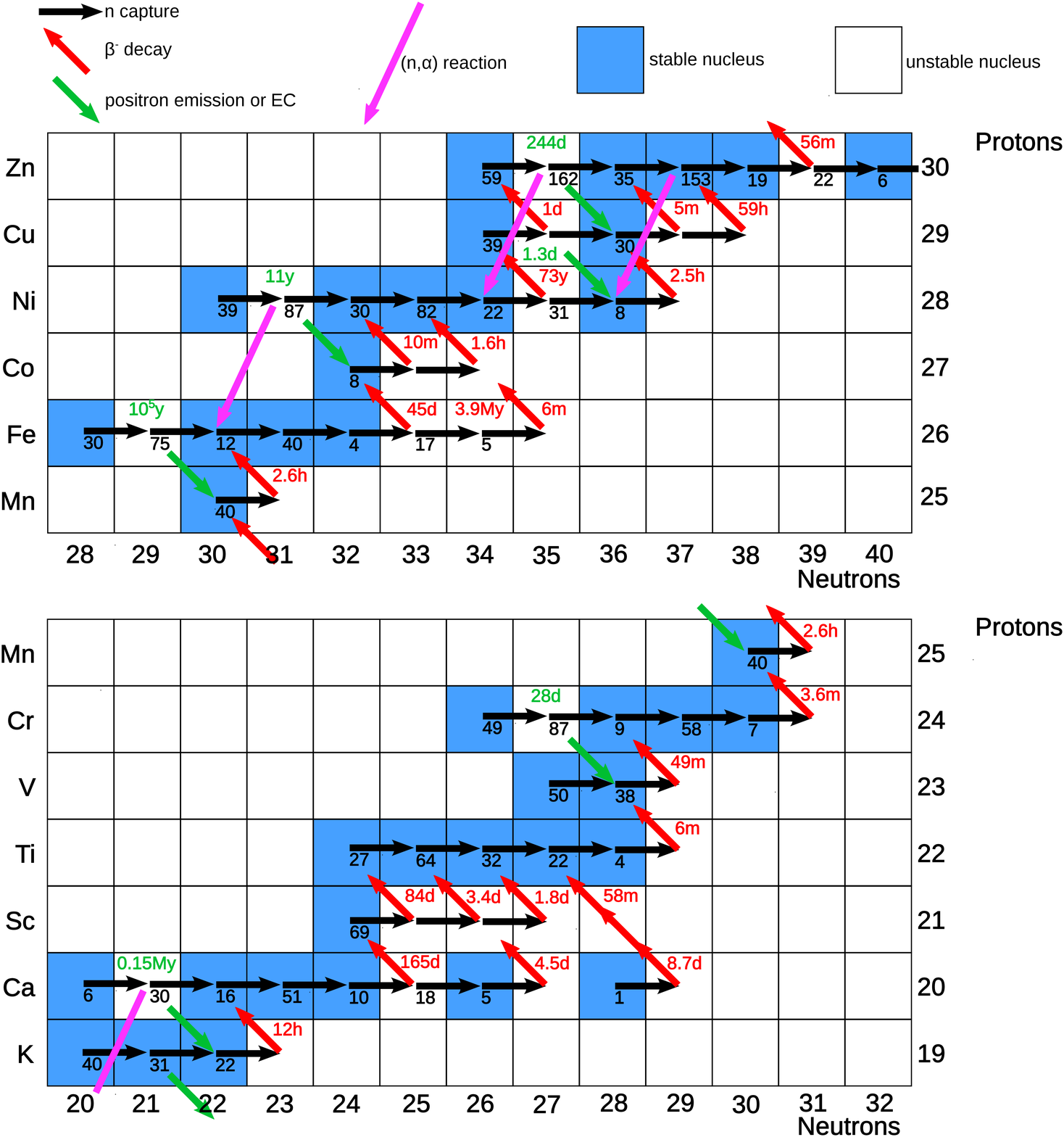}
\caption{\label{network}}
\end{figure*}

\vspace{1cm}
{{\bf Figure \ref{network}}. The network of reactions we considered to be active for the iron group nuclei relevant for our discussion. In the plot we show approximate 30keV values for cross sections (lower rows in the boxes indicating the nuclei) expressed in millibarns and laboratory lifetimes for unstable nuclei (upper rows). Note that these data are indicative: cross sections vary in a complex way as a function of temperature and also some lifetimes of radioactive species do the same. Note also that our network is, to our knowledge, exactly the same as adopted by the FRUITY database quoted in the text. Note also that our network is, to our knowledge, the same as adopted by the FRUITY database }.

\begin{figure*}
\includegraphics[width=0.5\textwidth]{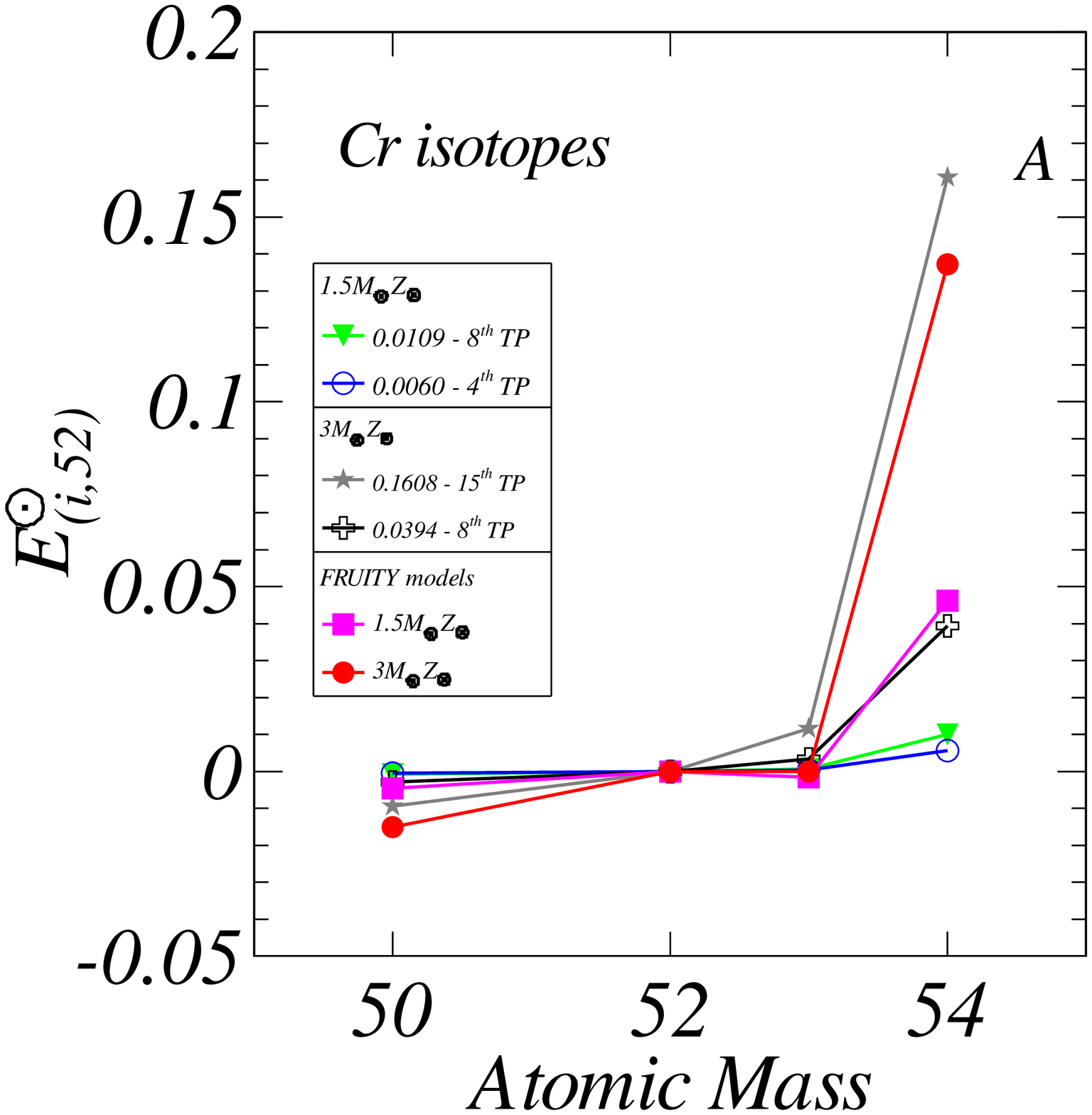}
\includegraphics[width=0.5\textwidth]{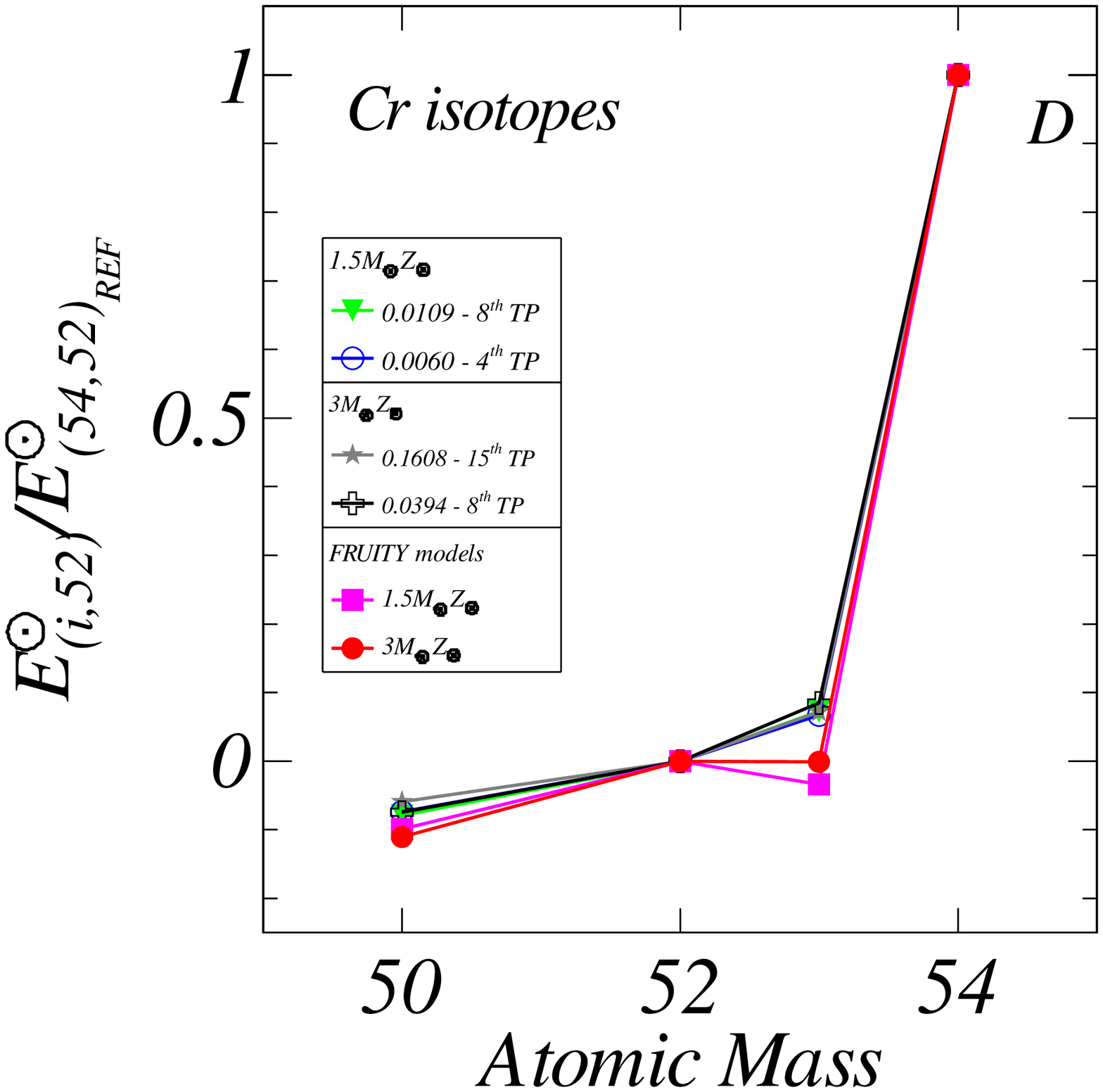}
\includegraphics[width=0.5\textwidth]{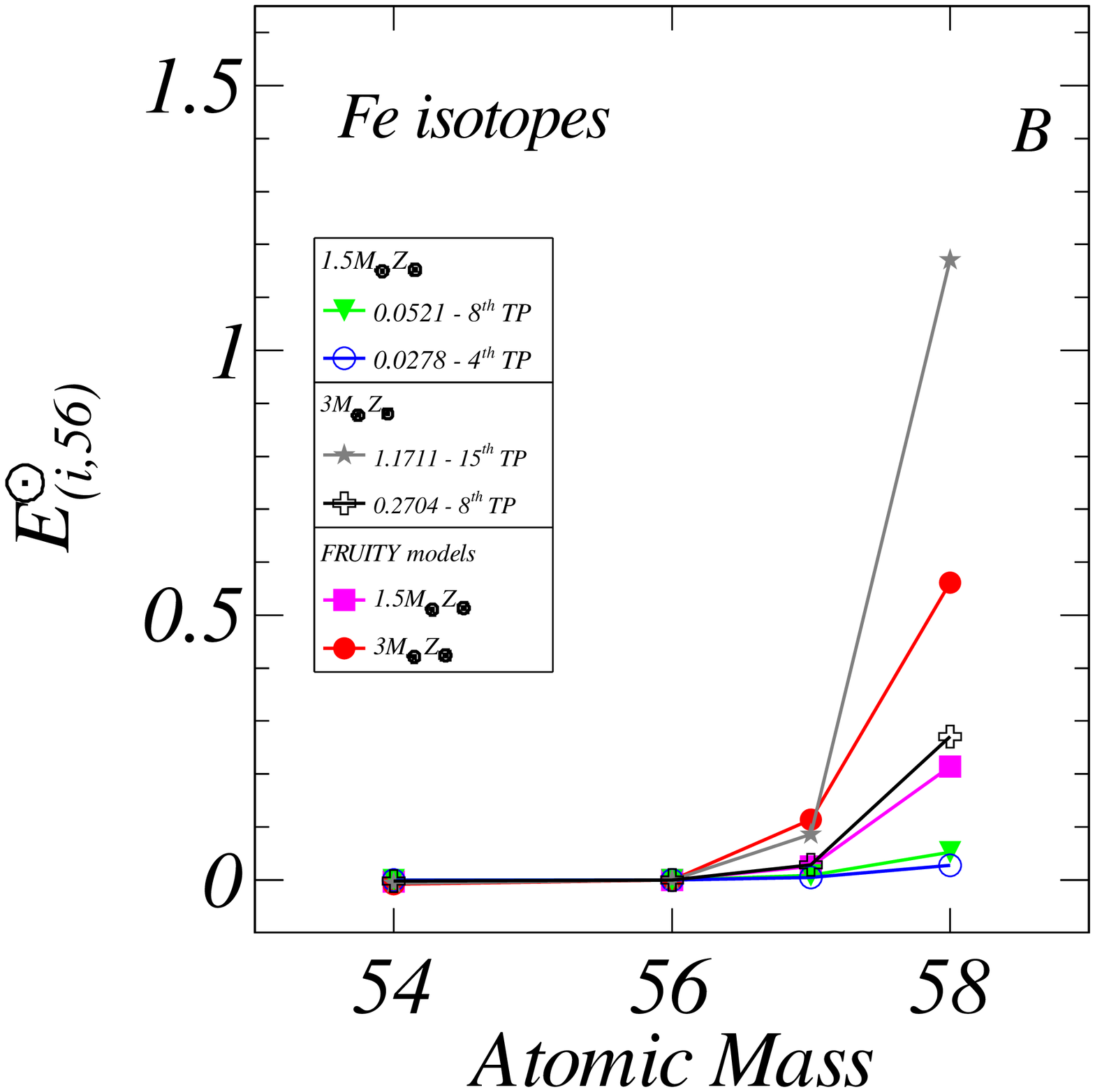}
\includegraphics[width=0.5\textwidth]{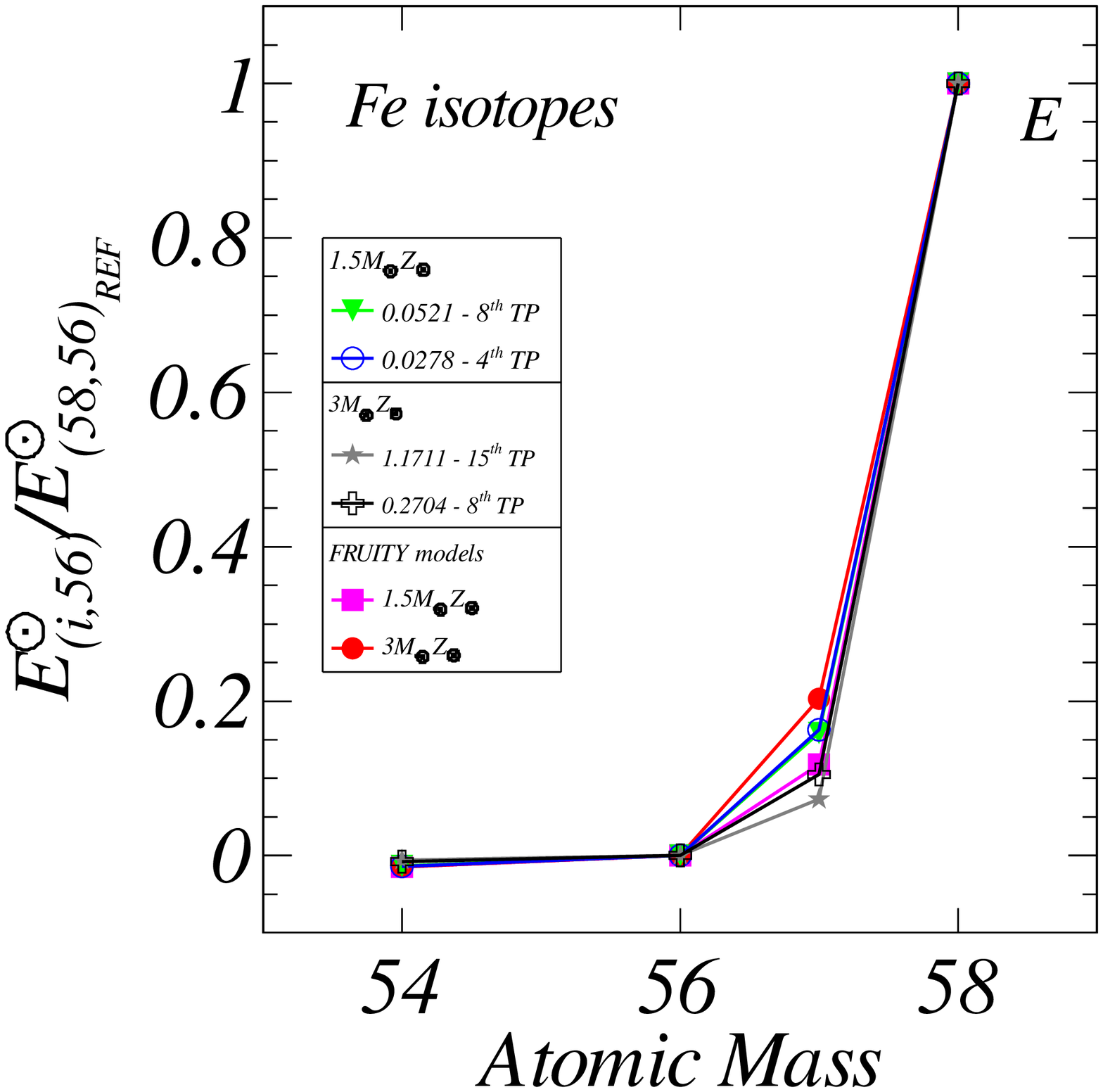}
\includegraphics[width=0.5\textwidth]{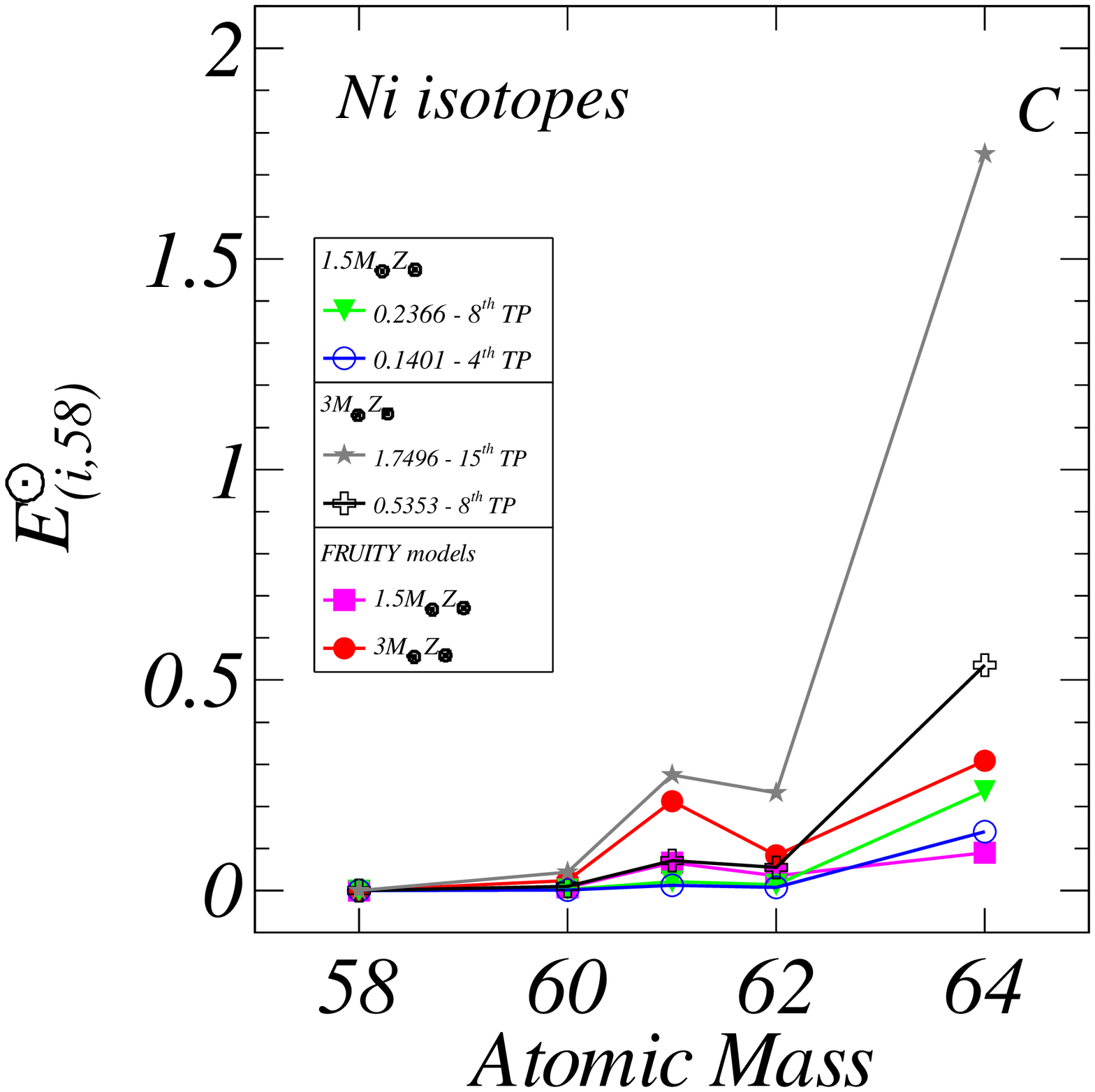}
\includegraphics[width=0.5\textwidth]{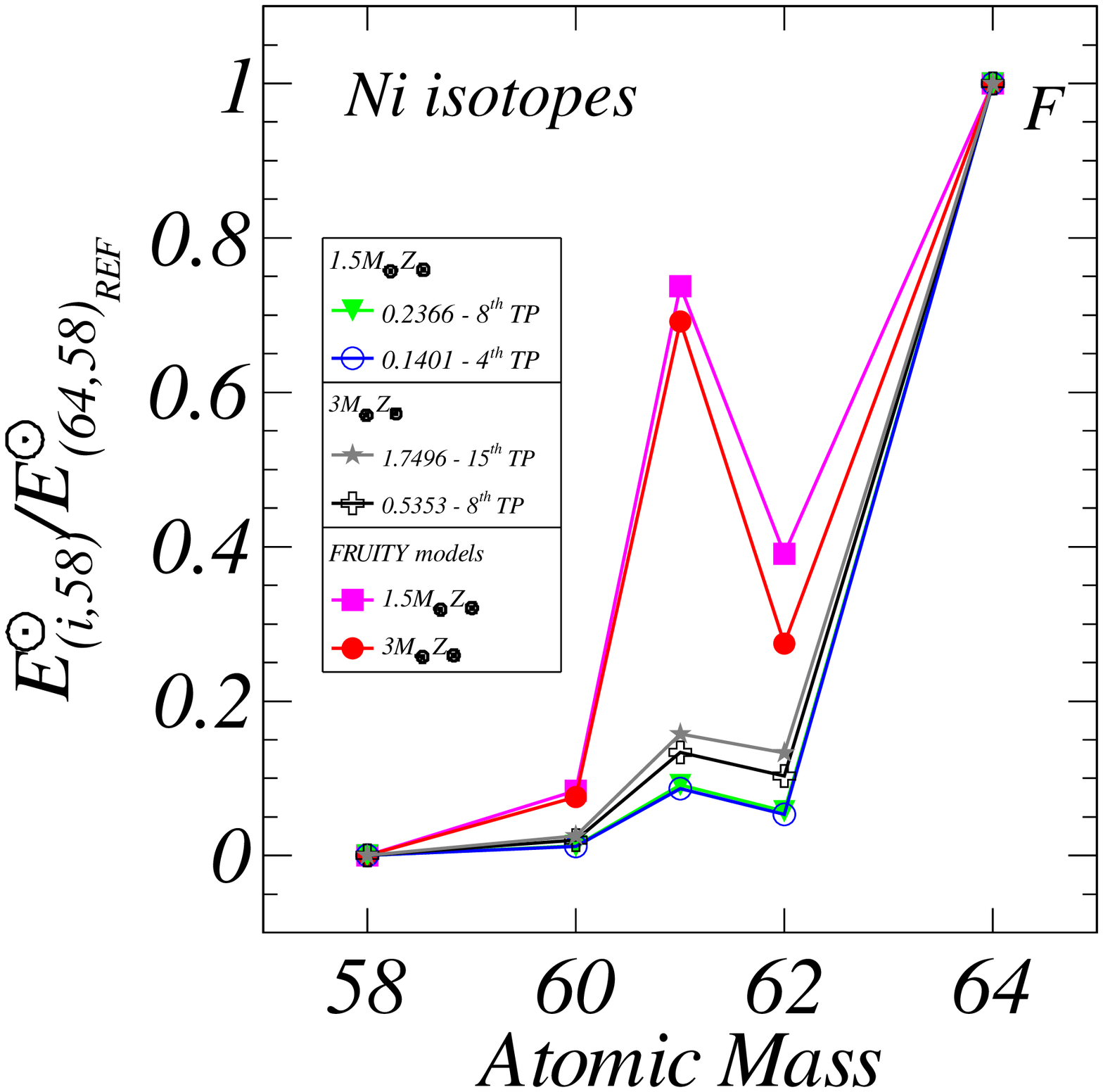}
\caption{\label{over1}}
\end{figure*}

\vspace{1cm}
{{\bf Figure \ref{over1}}. The enrichment factors ($E^{\odot}_{i,k}$) for Cr, Fe, and Ni for the envelopes of model AGB stars (with mass of $1.5$ and $3M_{\odot}$) having initially both a Solar metallicity, $Z = Z_{\odot}$, and Solar isotopic ratios, following the models by \citet{t+14}, but with the network updated for the Fe group elements as given in this study. Note that the calculation of $^{54}$Cr and $^{58}$Fe from our report lies well below that found in FRUITY for 1.5\ms. Figure \ref{over1} shows the cases for 1.5 and 3 \msb for different pulses and includes some results from the FRUITY on-line database \citet{c+11}, using their estimates for the enrichment factors in the envelope. The right part of the figure (Figure \ref{over1}, Panels d, e, and f) shows, for clear comparison, the enrichment factors for each case normalized to $E^{\odot}_{j,k}$ =1 where \oq $j$\cqb represents the isotope with the largest enhancement ($^{54}$Cr, $^{58}$Fe, and $^{64}$Ni). In our calculation the major shifts in isotopic abundance are clear. There is good agreement in the isotopic patterns for Cr and Fe for the results presented here and those from the FRUITY data base. There is a discrepancy in the case of Ni, due to the underproduction of $^{64}$Ni obtained from the FRUITY data base.}

\begin{figure*}
\includegraphics[width=0.5\textwidth]{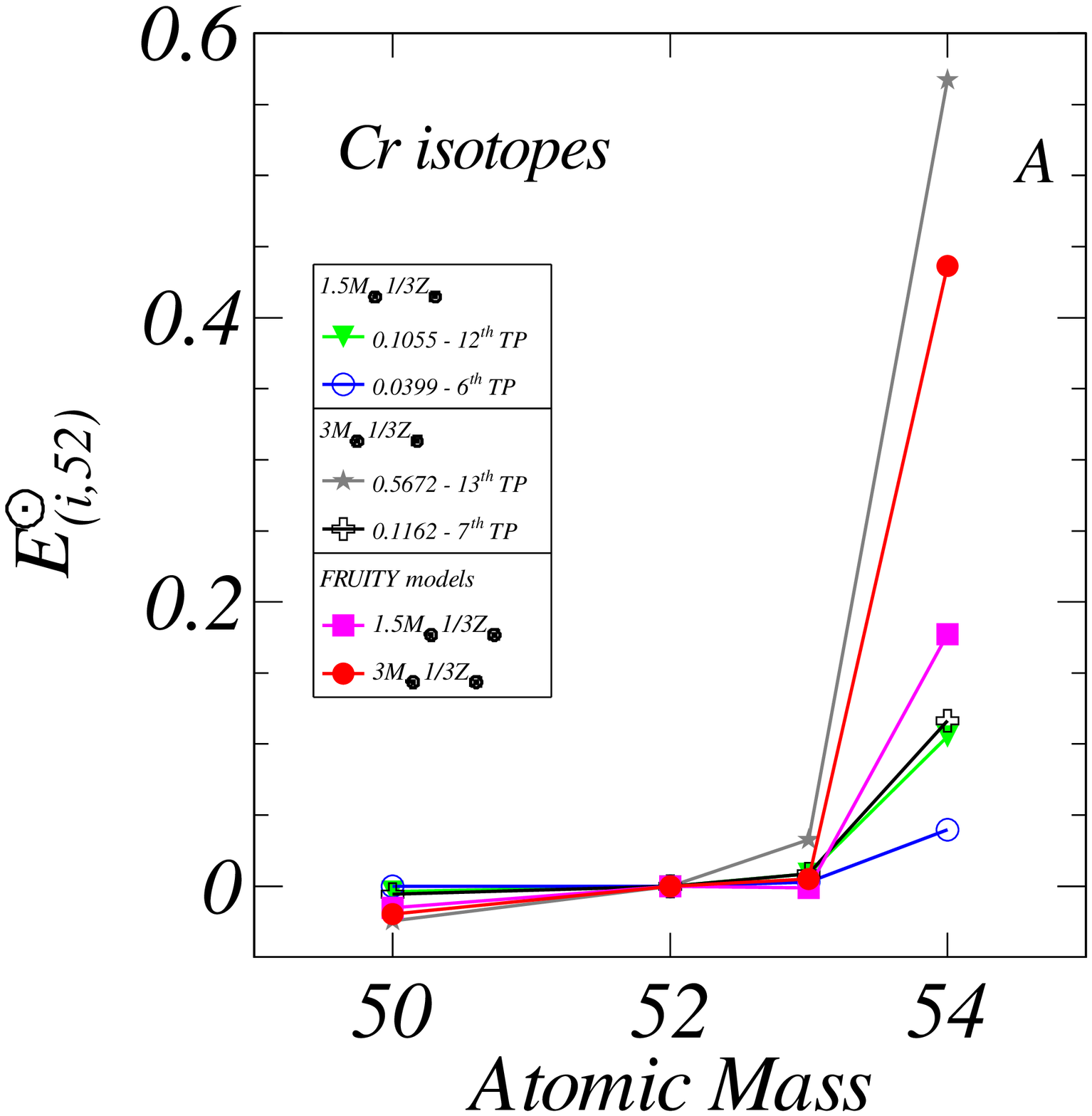}
\includegraphics[width=0.5\textwidth]{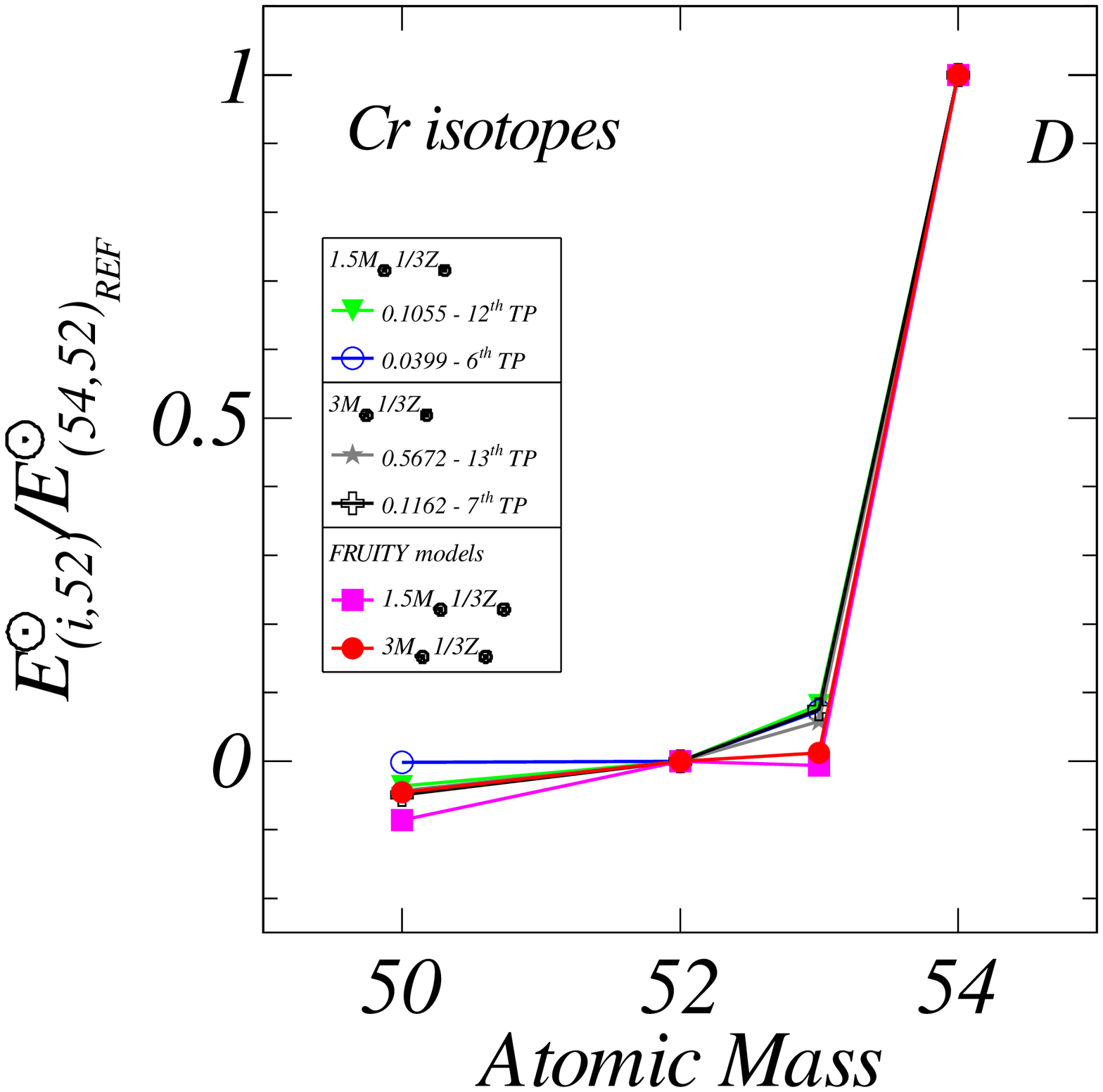}
\includegraphics[width=0.5\textwidth]{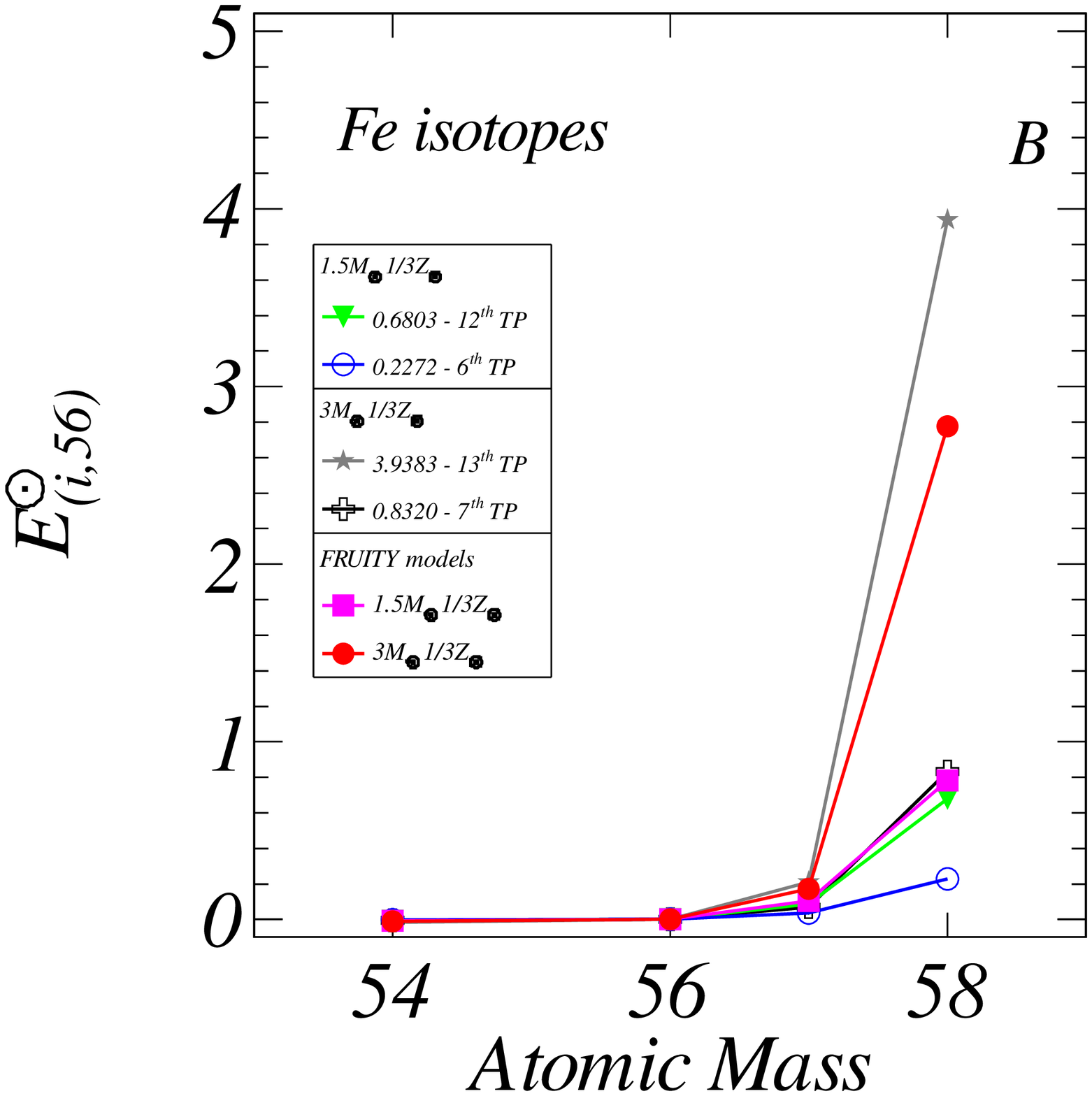}
\includegraphics[width=0.5\textwidth]{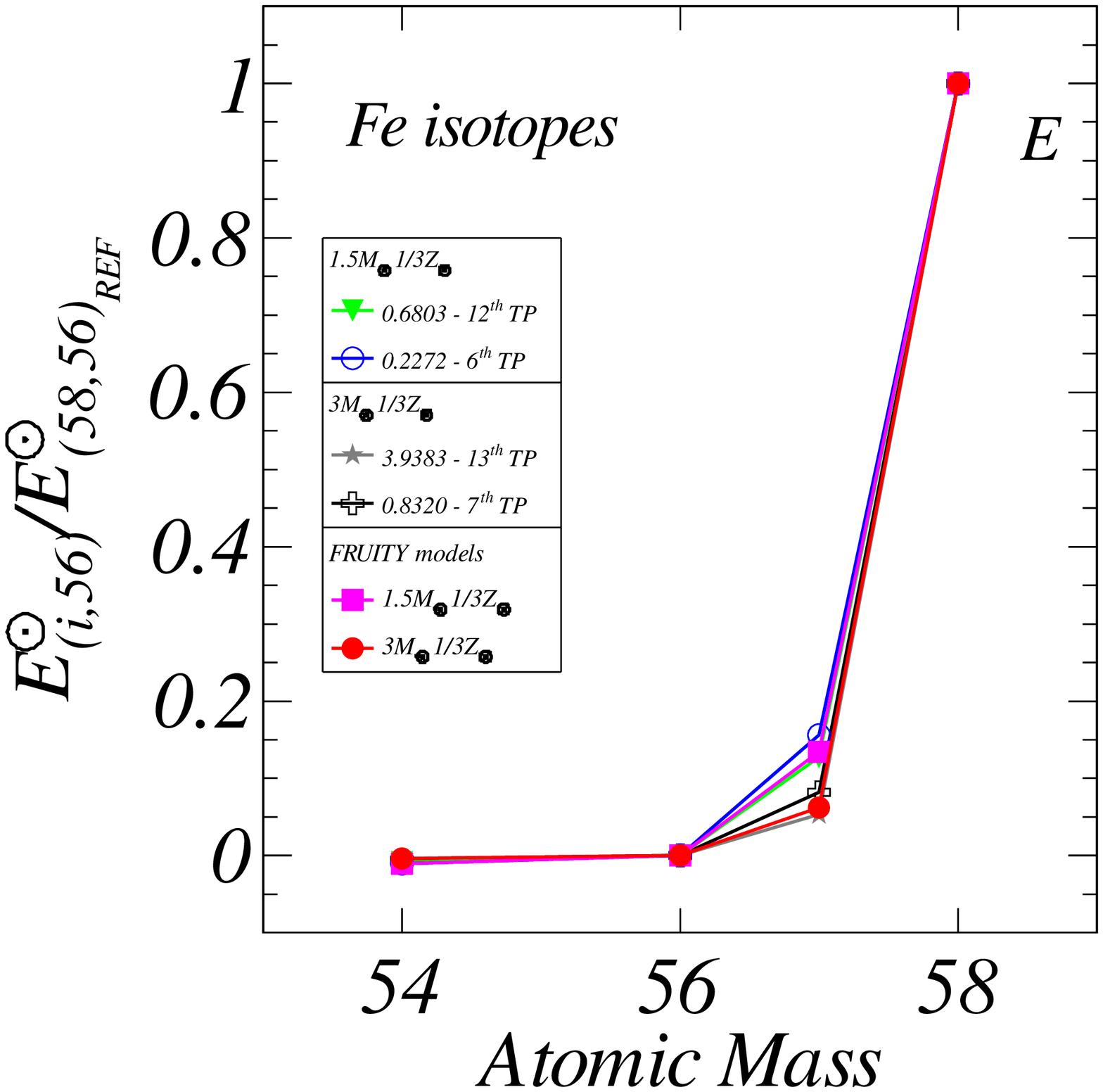}
\includegraphics[width=0.5\textwidth]{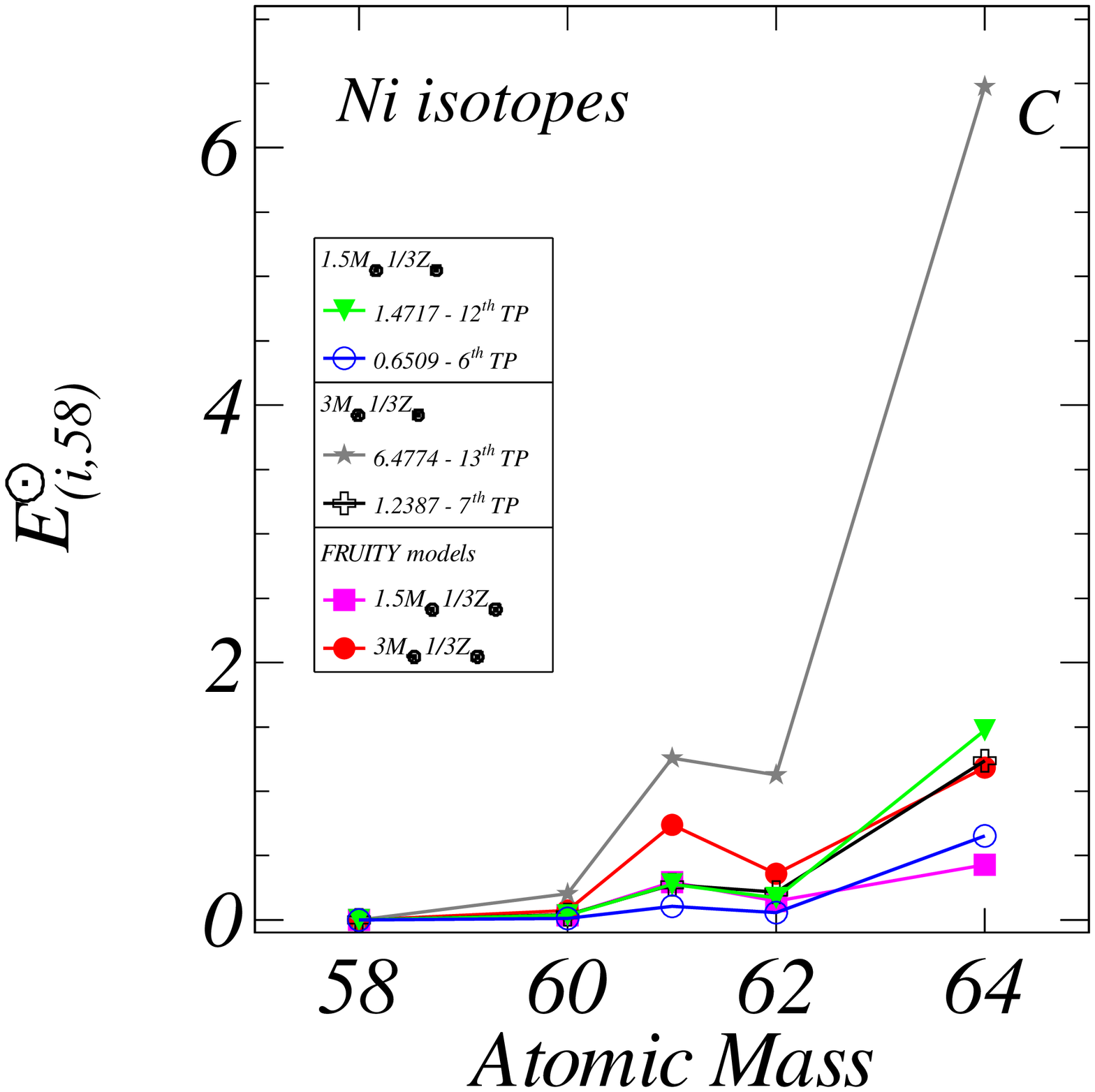}
\includegraphics[width=0.5\textwidth]{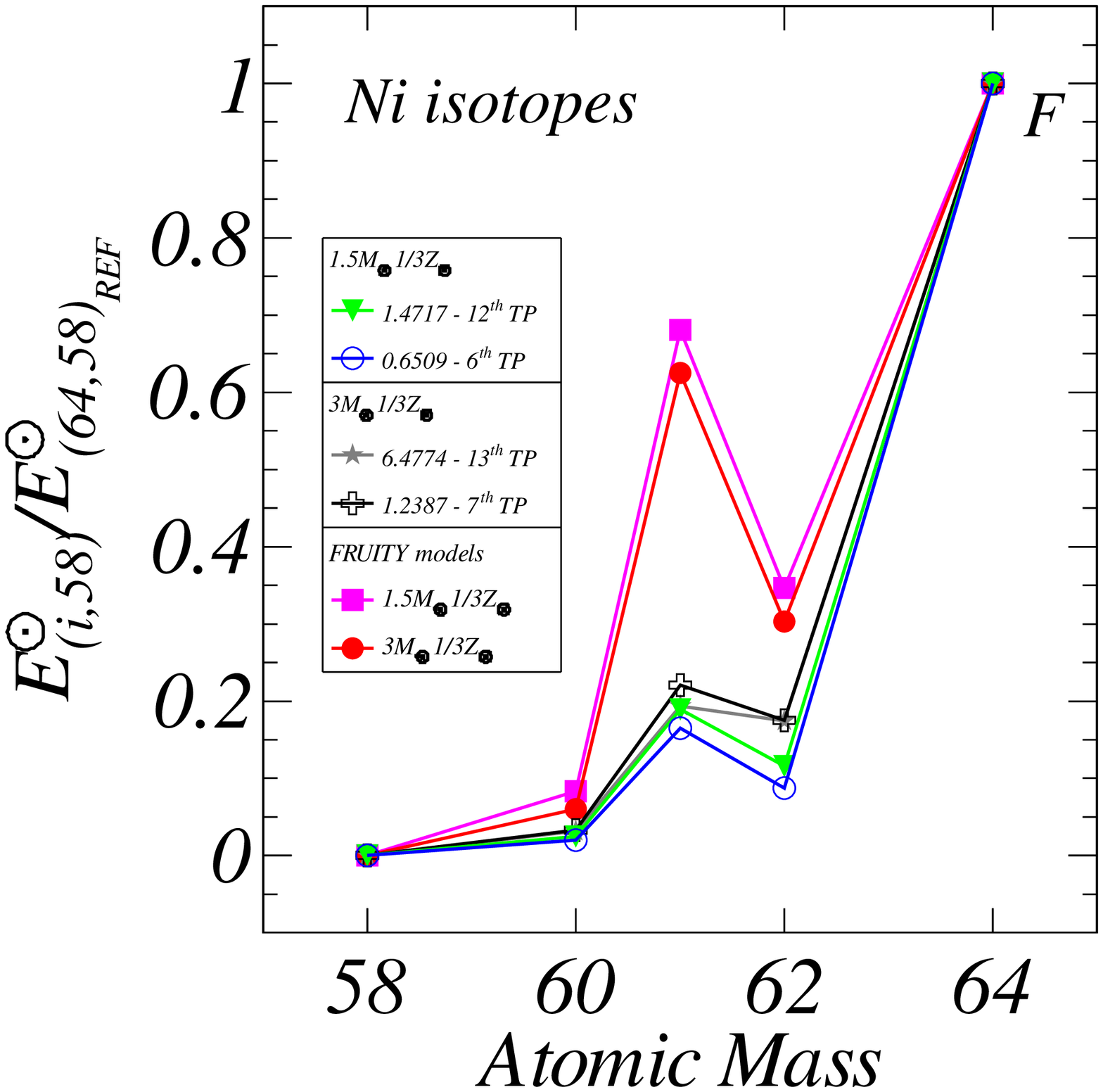}
\caption{\label{over2}}
\end{figure*}

\vspace{1cm}
{{\bf Figure \ref{over2}}. The enrichment factors ($E^{\odot}_{i,k}$) for the envelopes of model AGB stars with initial Solar isotopic ratios, but for $Z = Z_{\odot}/3$. See details in the caption of Figure \ref{over1}. Note that here, the data points for $^{58}$Fe and $^{54}$Cr in the 1.5\msb case by both calculations are close.}

\begin{figure*}
\includegraphics[width=\textwidth]{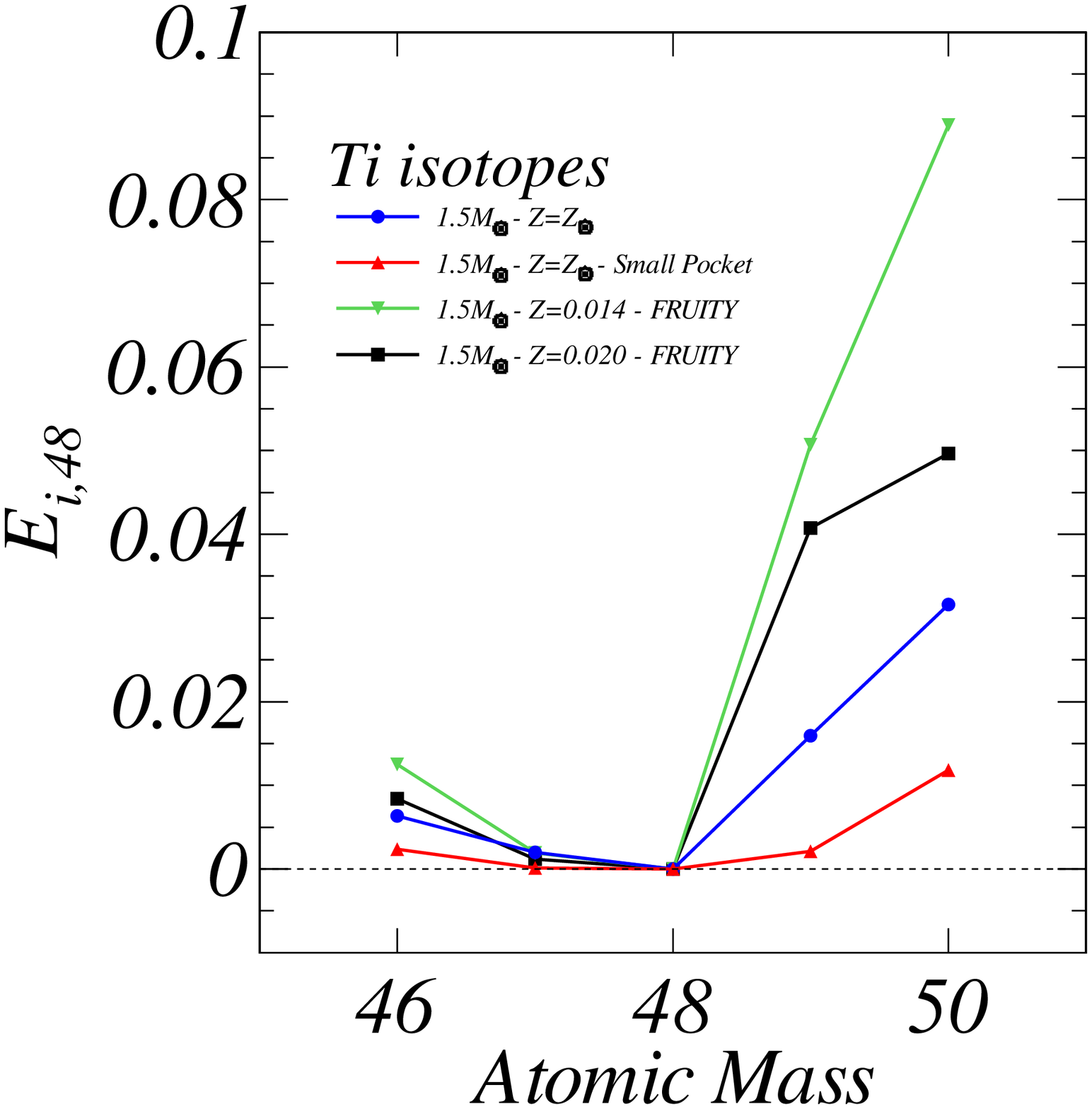}
\caption{\label{tifr}}
\end{figure*}

\vspace{1cm}
{\bf Figure \ref{tifr}}. The $E_{i,k}$-values for Ti at the last pulse from the FRUITY outputs, for two values of the metallicity that bracket our choice ($Z_{\odot} = 0.0154$). As a comparison, the results we obtained in this paper, both for our standard choice of
the $^{13}$C pocket mass \citep[see Figure 1 in][]{t+14} and for the smaller pocket of 0.001 \ms. The isotopic patterns for both calculations are essentially the same. However the much larger effects for the Fruity output as compared to 
our calculations must be investigated.

\begin{figure*}
\includegraphics[width=0.5\textwidth]{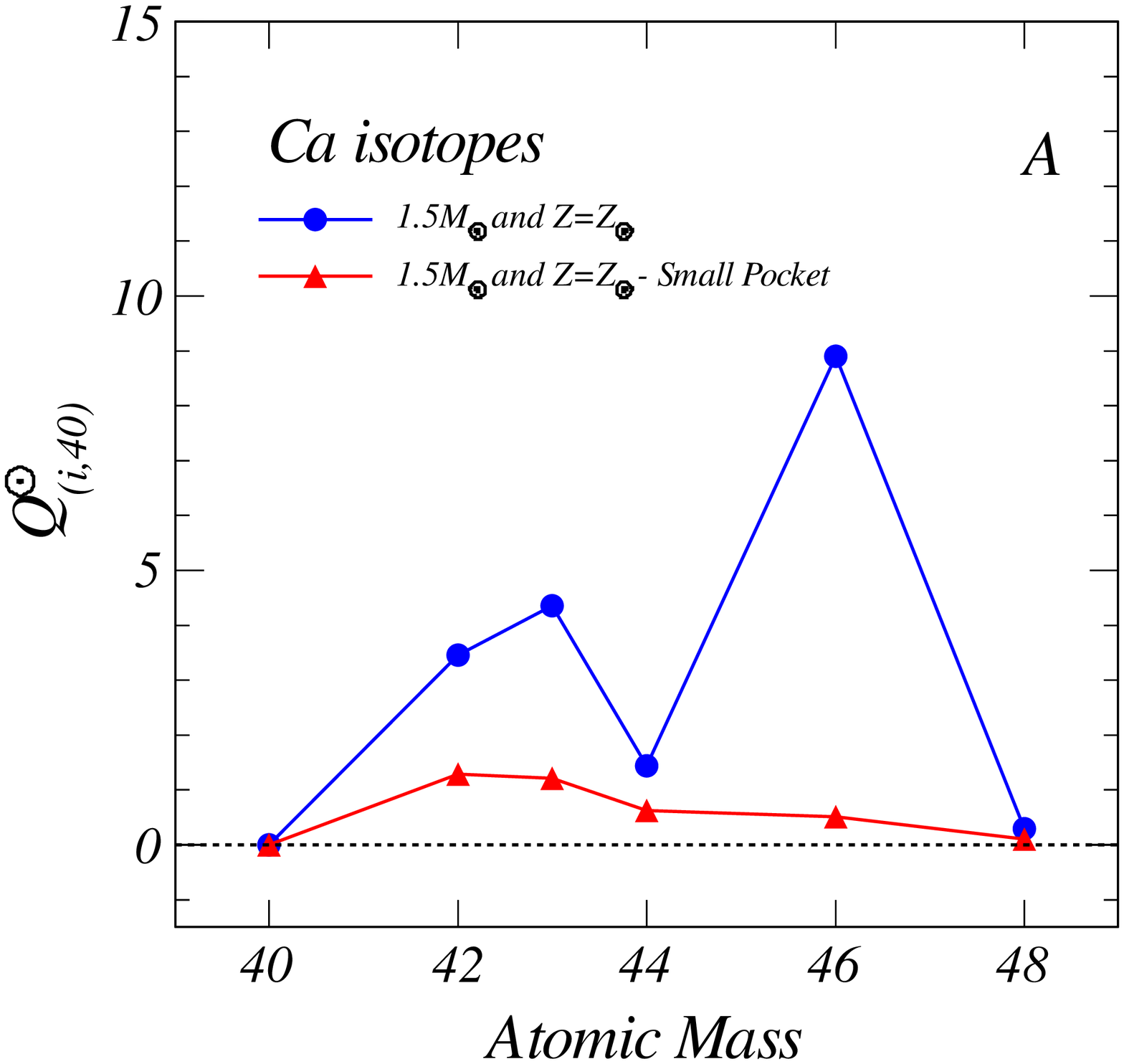}
\includegraphics[width=0.5\textwidth]{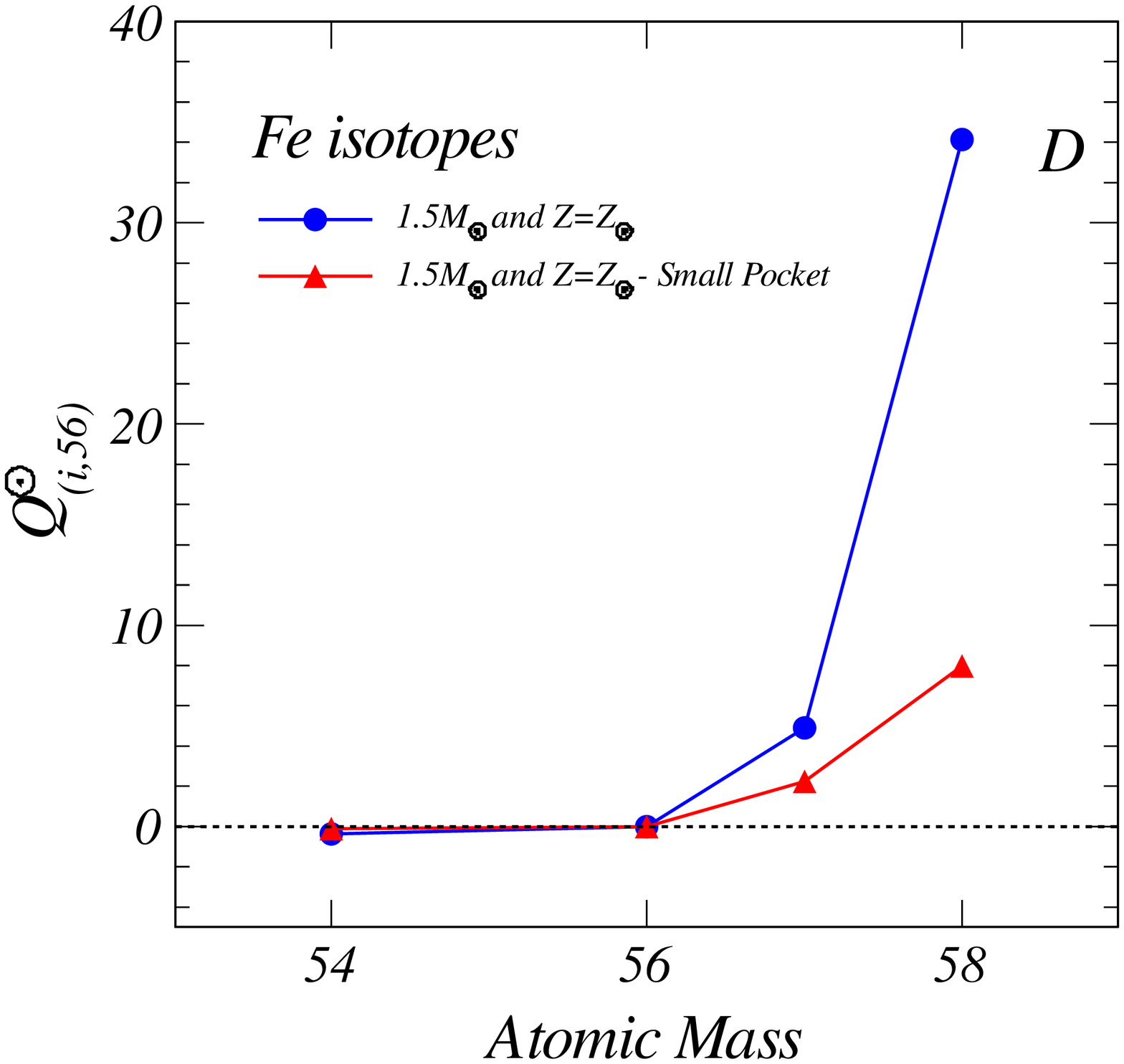}
\includegraphics[width=0.5\textwidth]{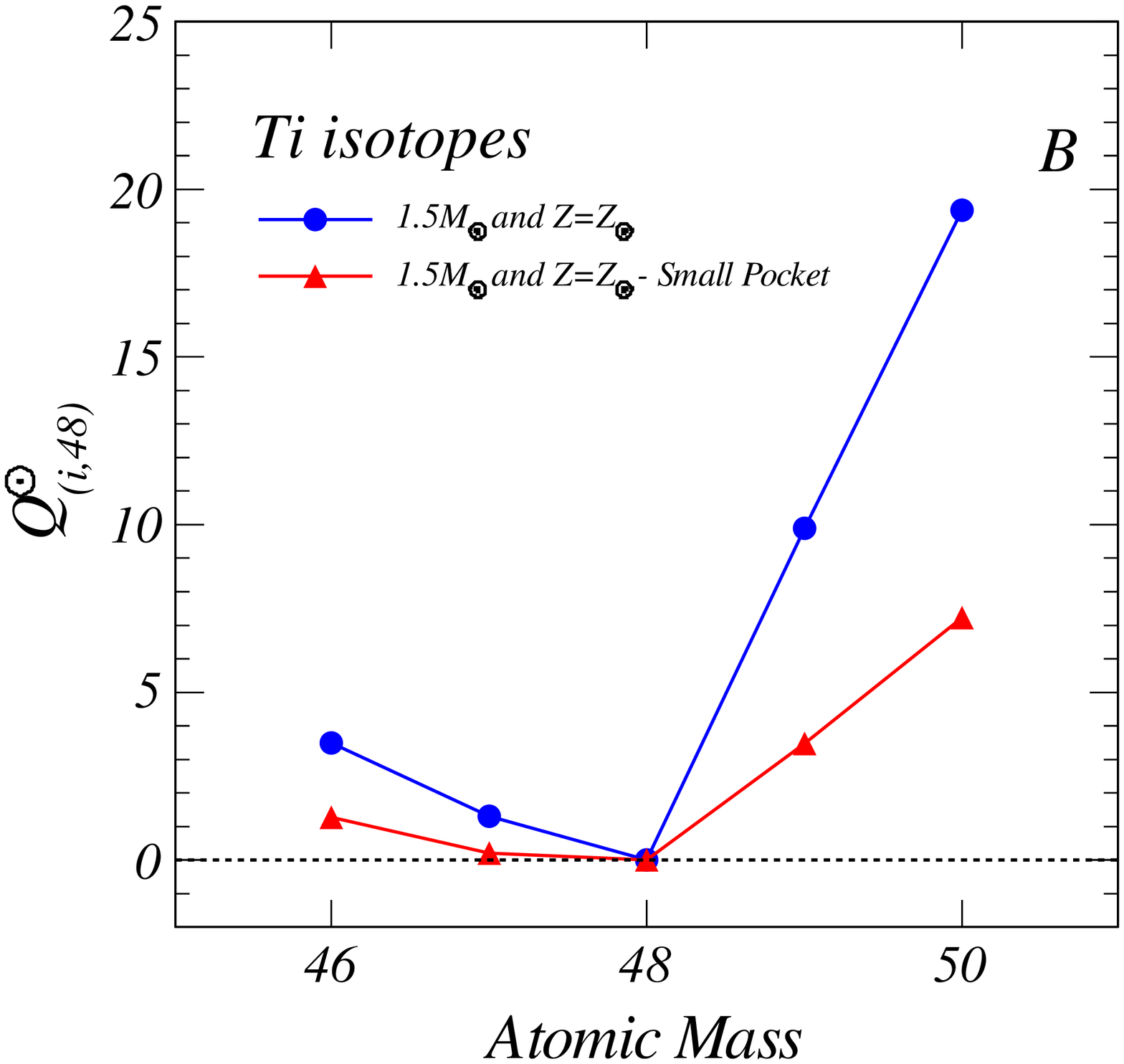}
\includegraphics[width=0.5\textwidth]{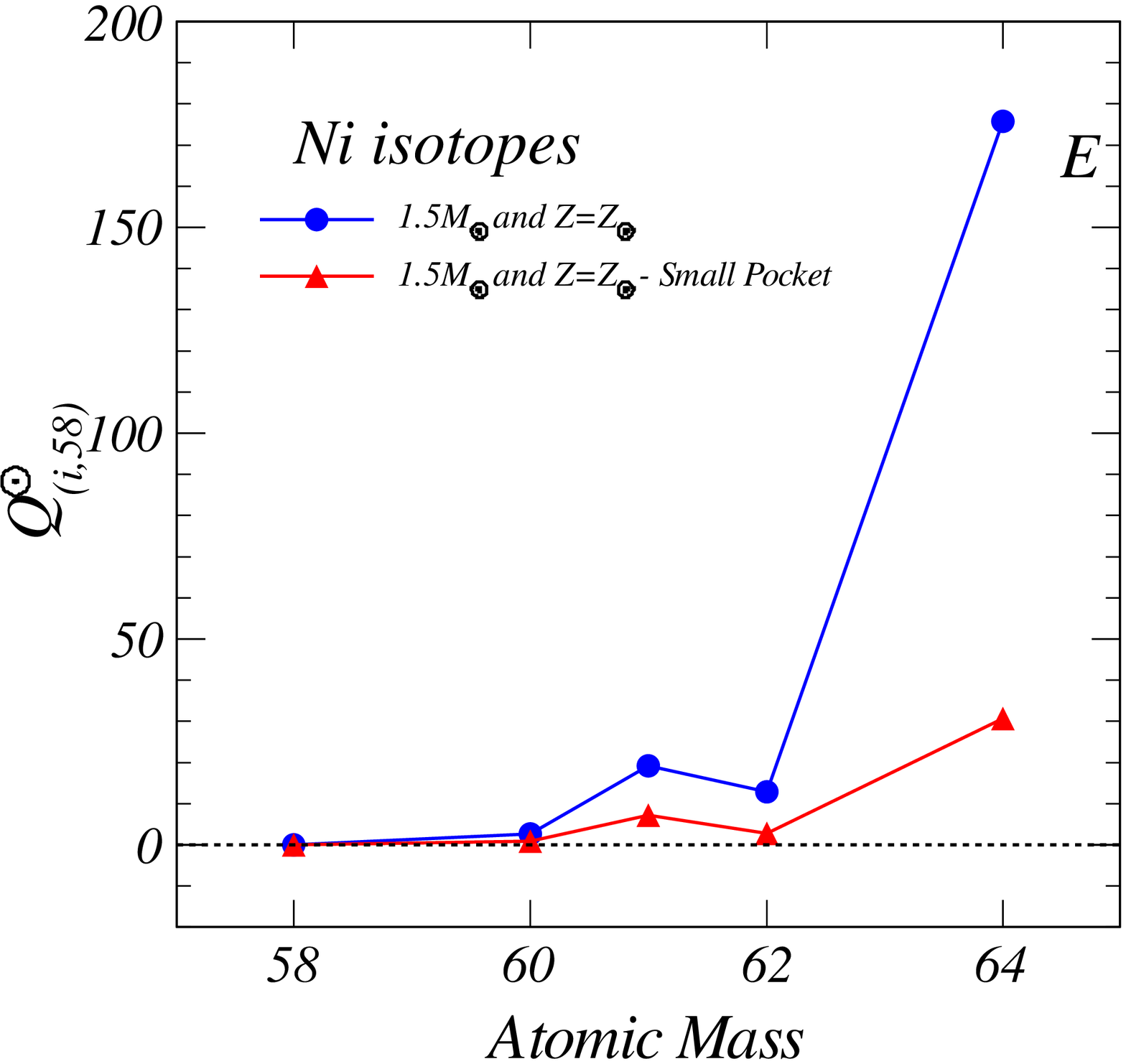}
\includegraphics[width=0.5\textwidth]{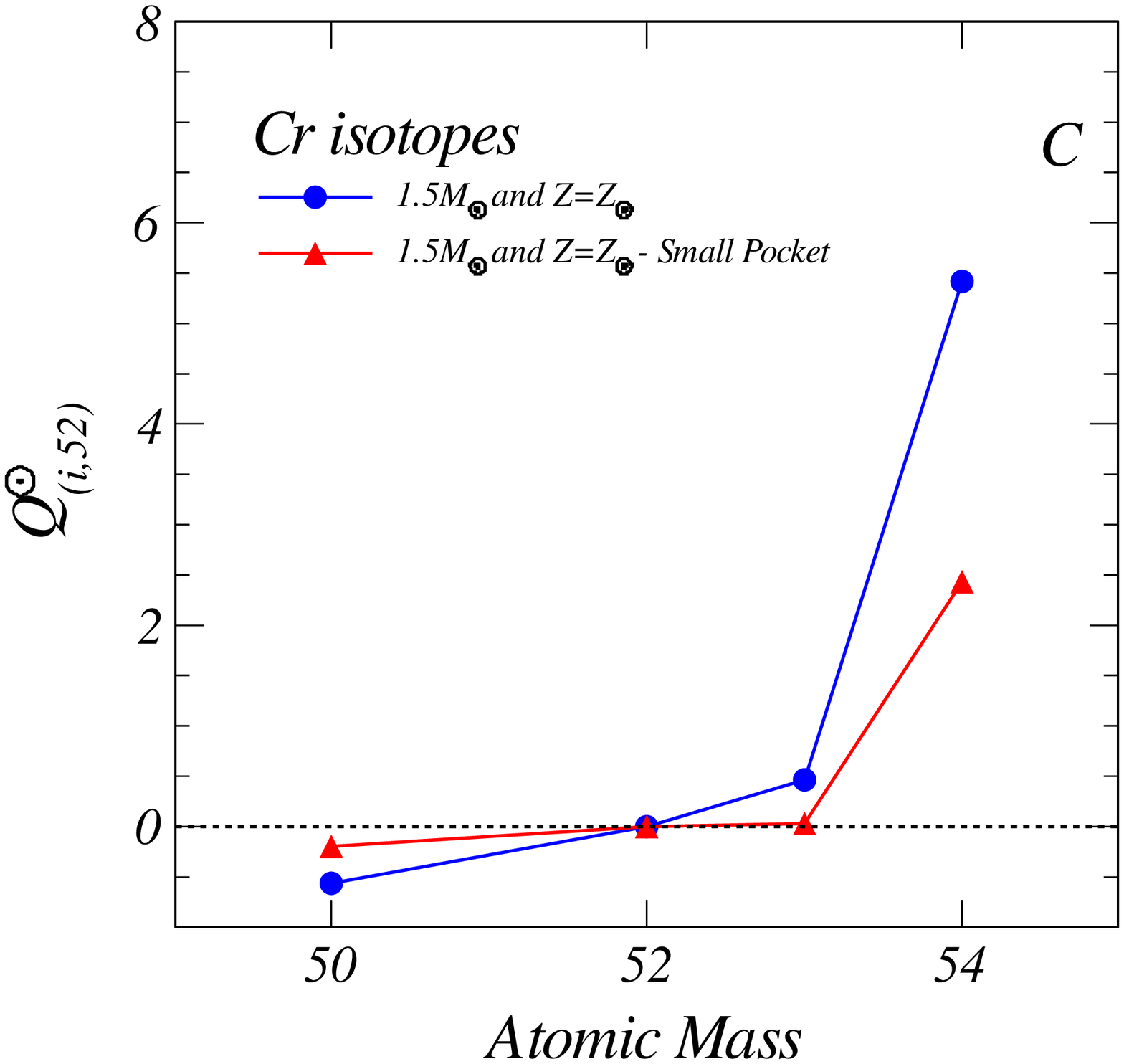}
\includegraphics[width=0.5\textwidth]{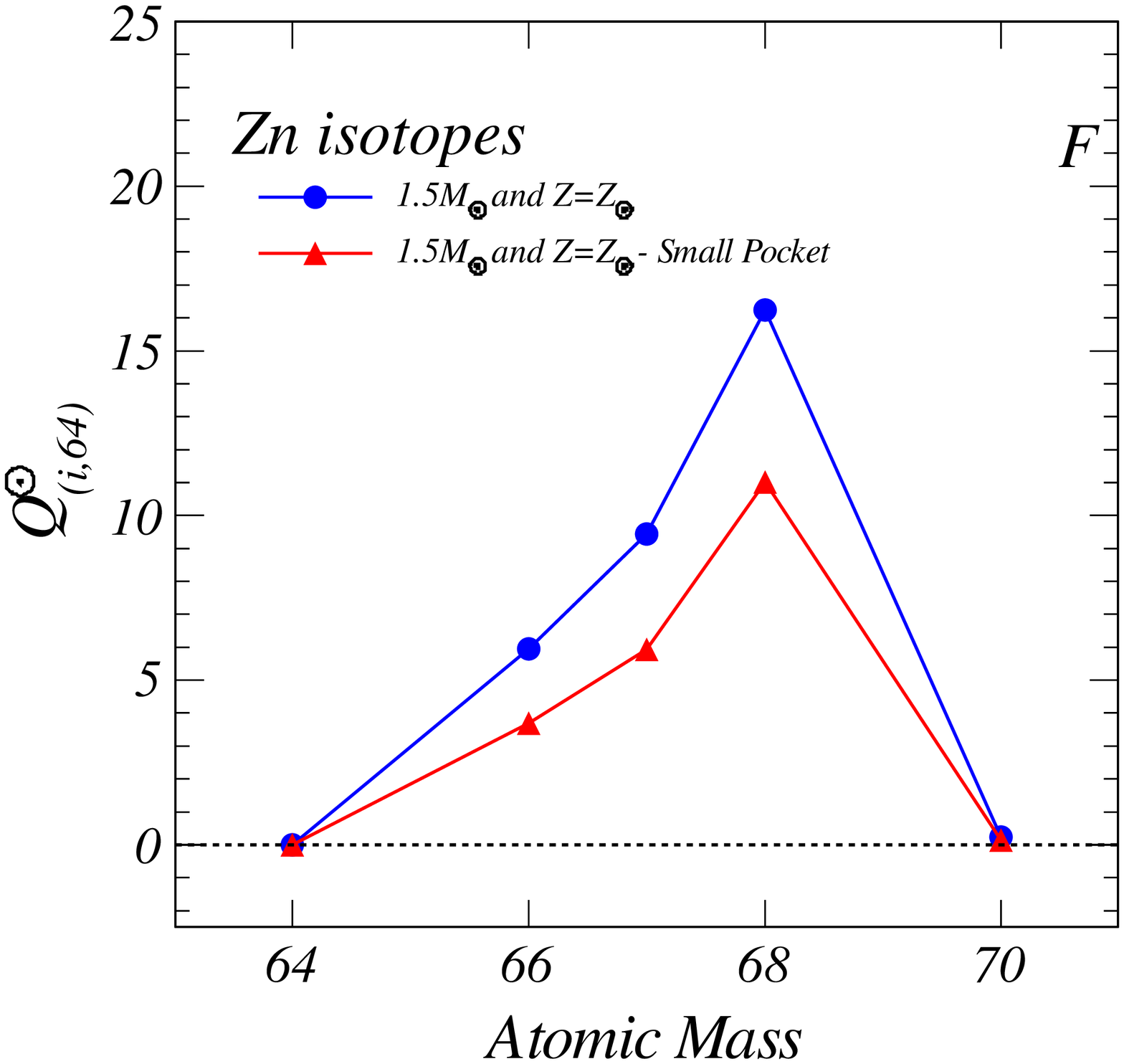}
\caption{\label{q_values}}
\end{figure*}

\begin{figure*}
\includegraphics[width=0.5\textwidth]{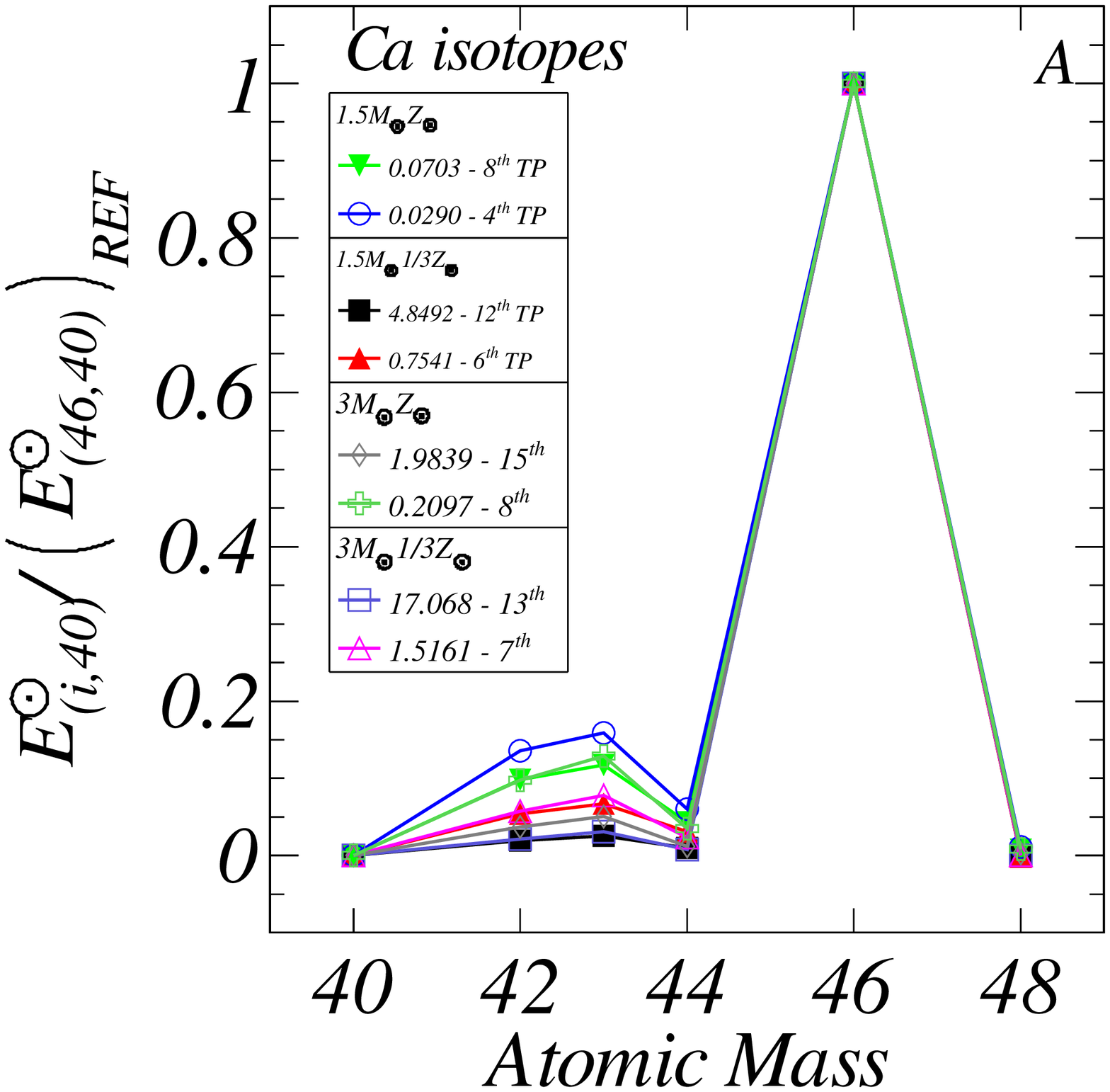}
\includegraphics[width=0.5\textwidth]{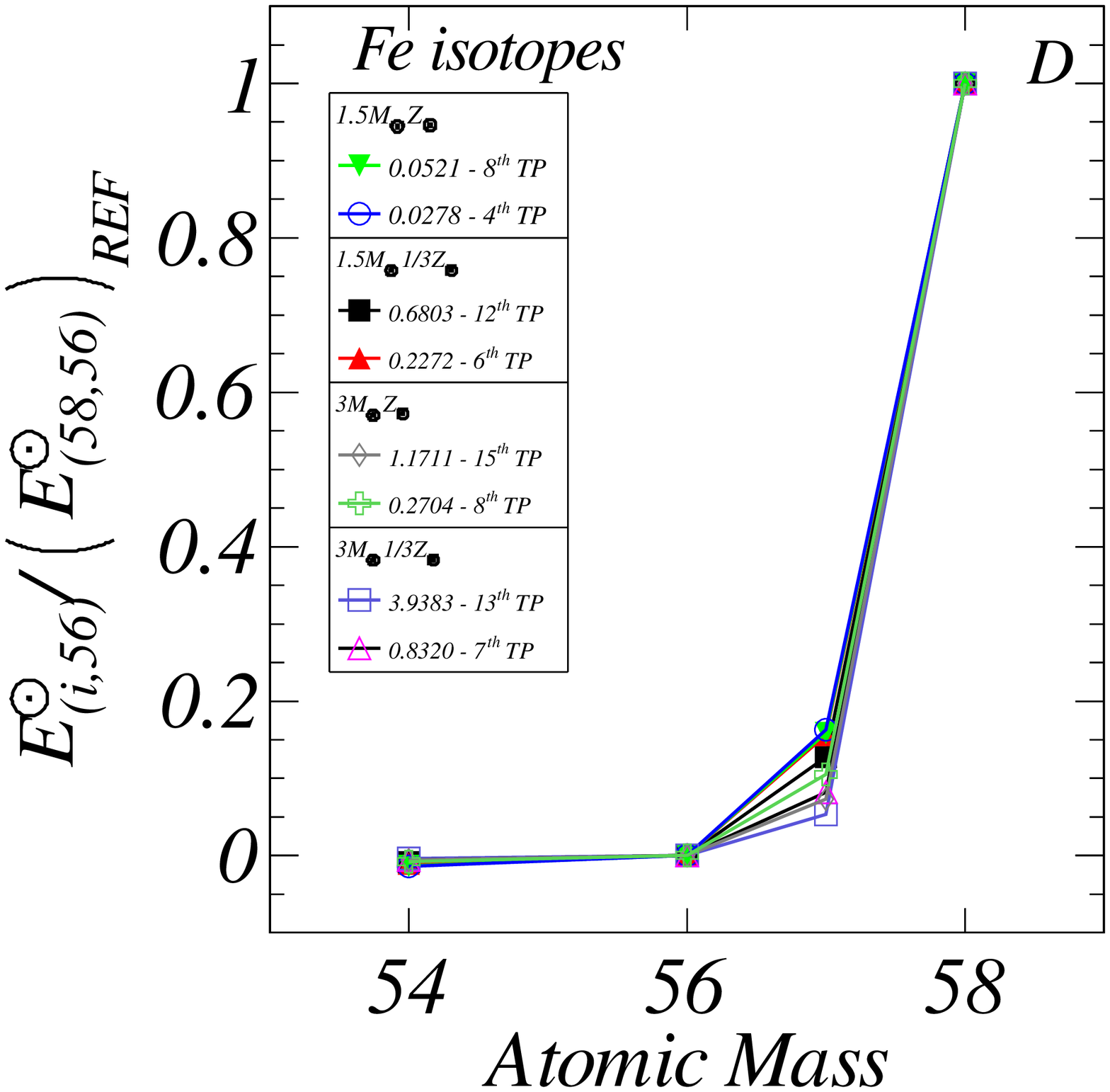}
\includegraphics[width=0.5\textwidth]{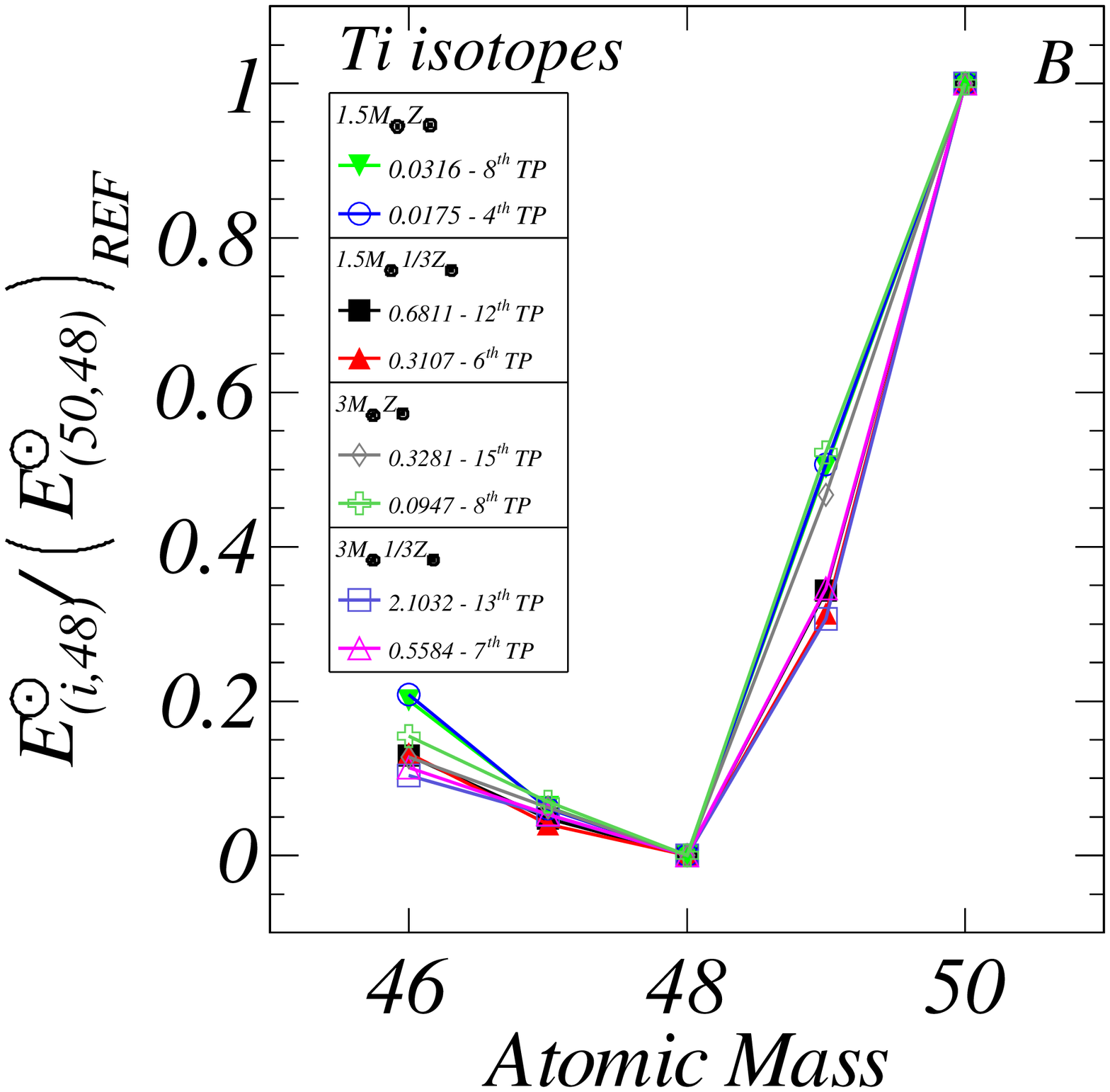}
\includegraphics[width=0.5\textwidth]{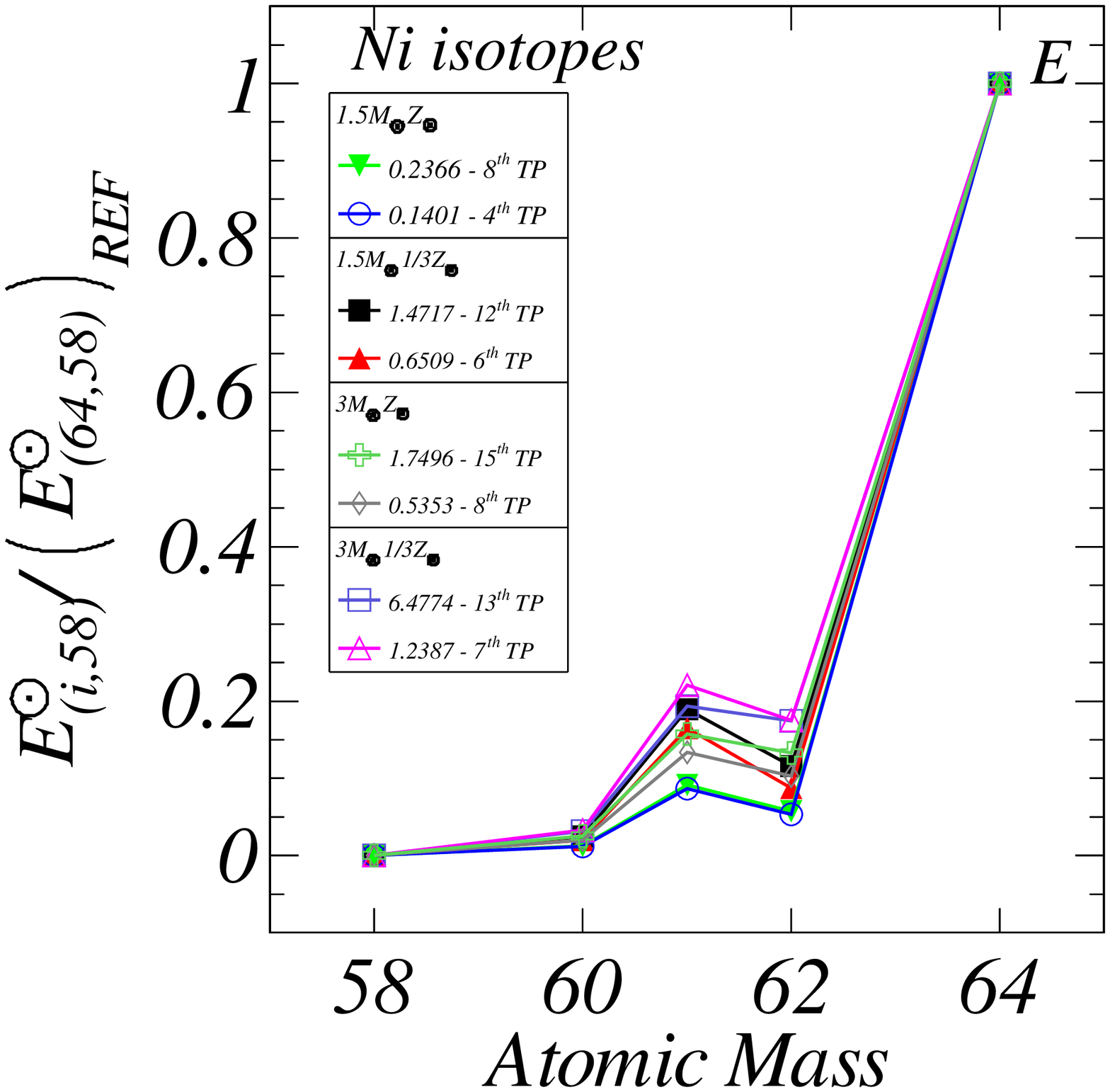}
\includegraphics[width=0.5\textwidth]{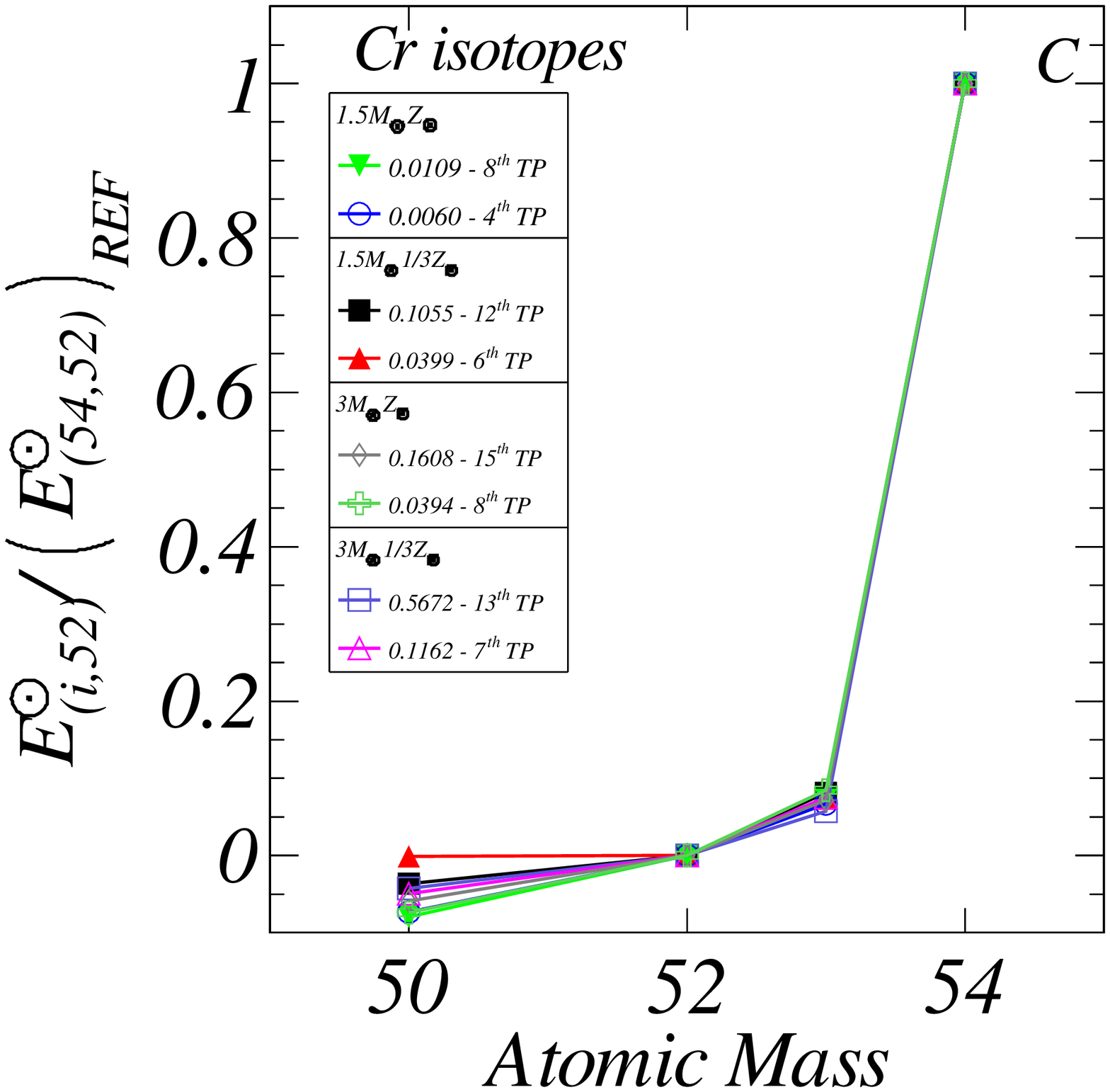}
\includegraphics[width=0.5\textwidth]{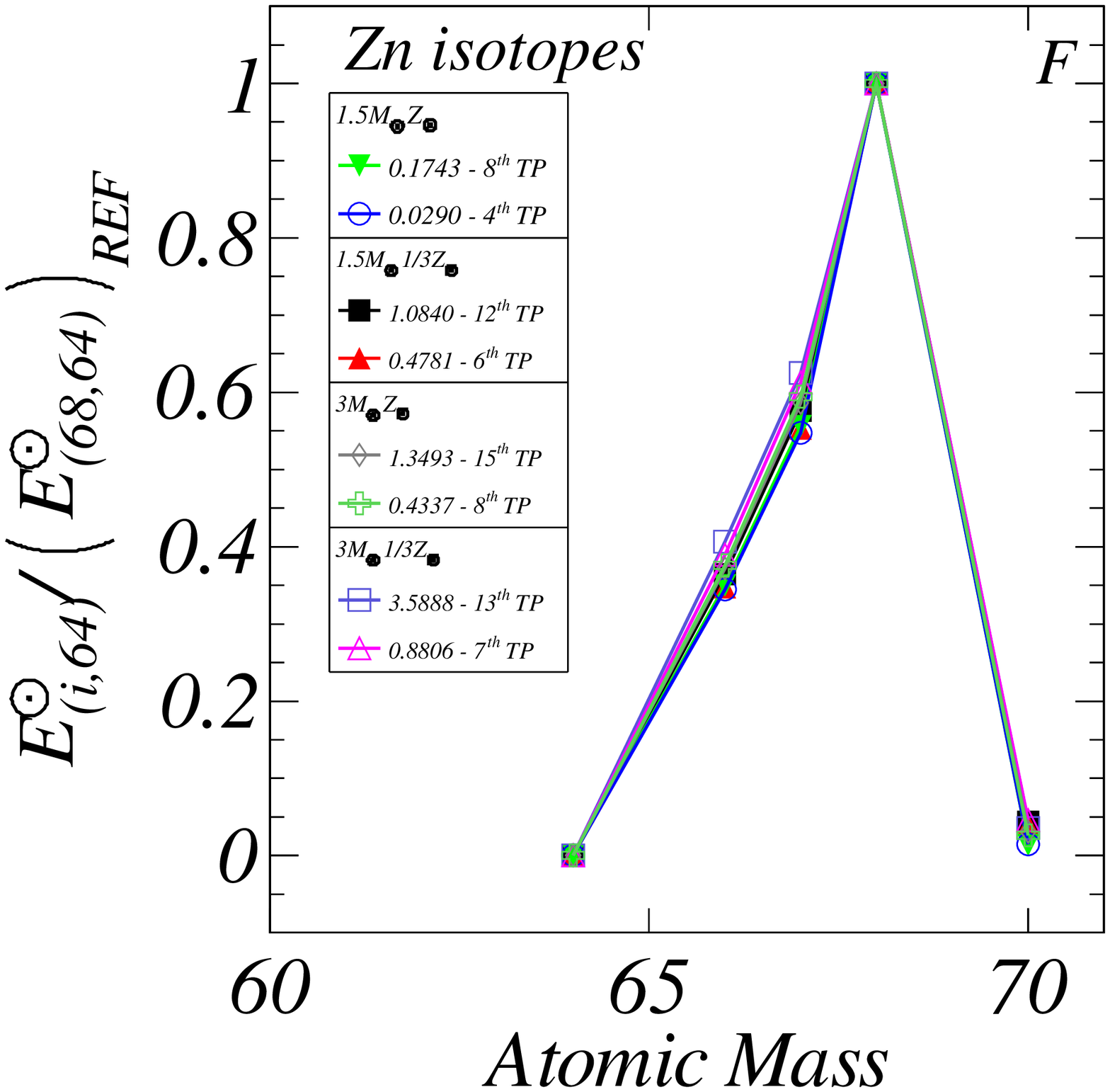}
\caption{\label{over}}
\end{figure*}

\vspace{1cm}
{{\bf Figure \ref{q_values}}. Isotopic enrichments for a wide suite of elements of $s$-processed matter in a parcel before mixing in to the envelope $Q^{\odot}_{i,k}$ (see section 2.3.1 in text). The values of $Q^{\odot}_{i,k}$ are arbitrarily chosen here at the last pulse in a 1.5 \msb model. Solar initial isotopic values were used in all these calculations. The index isotopes \oq $k$\cqb used are $^{40}$Ca, $^{48}$Ti, $^{52}$Cr, $^{56}$Fe, $^{58}$Ni, and $^{64}$Zn. Comparison should be made between the patters represented by $Q_{i,k}$ and the envelope enrichment patterns shown in figure \ref{over}. It is important to recognise that to reach $E_{50,48}$ around unity would require, from our calculations, a more massive star and low metallicity ($Z_{\odot}$/3) source (see numerical numbers in Figure \ref{over}). Full red triangles represent the $Q$ values for smaller $^{13}$C pocket \citep[in order to reproduce the effects suggested by e.g.][]{g+98,was06} from the Case A calculations of \citet{t+14}.}

\vspace{1cm}
{{\bf Figure \ref{over}}. Summary of the isotopic compositions for a wide variety of \oq Fe group\cqb elements in the envelope of AGB stars. We have represented the results normalized to $E^{\odot}_{j,k}$ =1 for those isotopes \oq $j$\cqb that are most enhanced by $s$-processing in our calculations. These are, respectively, $^{46}$Ca, $^{50}$Ti, $^{54}$Cr, $^{58}$Fe, $^{64}$Ni, and $^{68}$Zn. The index isotopes \oq $k$\cqb used are $^{40}$Ca, $^{48}$Ti, $^{52}$Cr, $^{56}$Fe, $^{58}$Ni, and $^{64}$Zn. Notice that the form of these curves is almost independent of the pulse, the stellar mass or the metallicity. The curves shown here should be compared with the $Q^{\odot}_{i,k}$ shown in Figure \ref{q_values}. Ca shows the most discrepant behaviour. Otherwise, the curves are almost indistinguishable. See text for discussion of the individual elements. Note the peak at $^{46}$Ca, resulting from using the laboratory lifetime of $^{45}$Ca. This effect is not supported by observations.}

\begin{figure*}
\includegraphics[width=\textwidth]{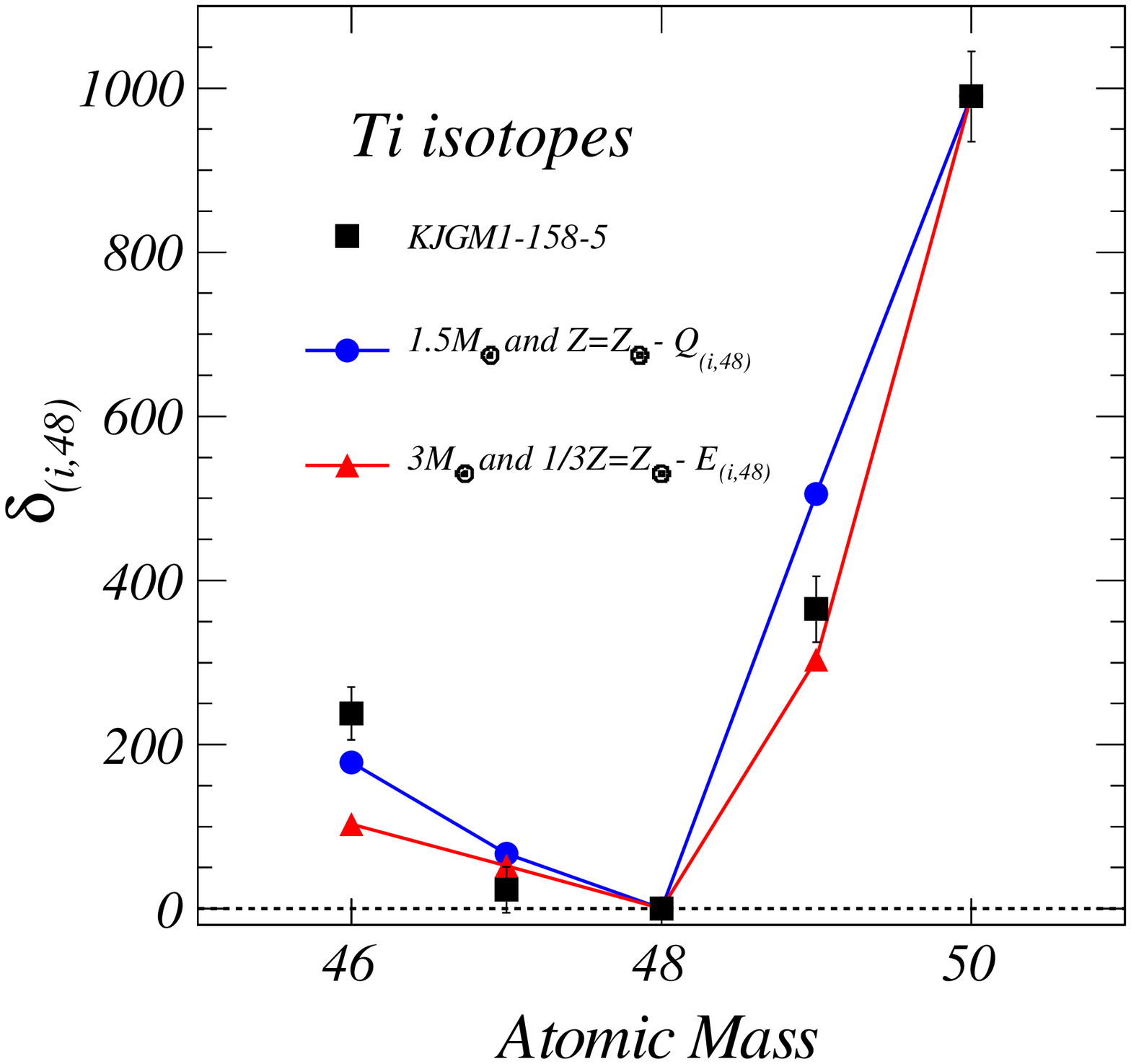}
\caption{\label{ti-1}}
\end{figure*}

\vspace{1cm}
{{\bf Figure \ref{ti-1}}. Measurements on Ti with quoted errors \citep{amari1} for the CIRCON grain KJGM1-158-5, which has the largest measured enrichment in $^{50}$Ti. For comparison, we show the curve for $Q^{\odot}_{i,48}$ in star a of 1.5\msb and $Z_{\odot}$, which is normalized to $\delta_{50,48}$ = 990 per mil. Note the very good agreement between these data and AGB $s$-processing. It's important to recognize that to reach $E_{50,48}$ around unity would require, from our calculations, a more massive (3\ms) star and low metallicity ($Z_{\odot}$/3) source.}

\begin{figure*}
\includegraphics[width=\textwidth]{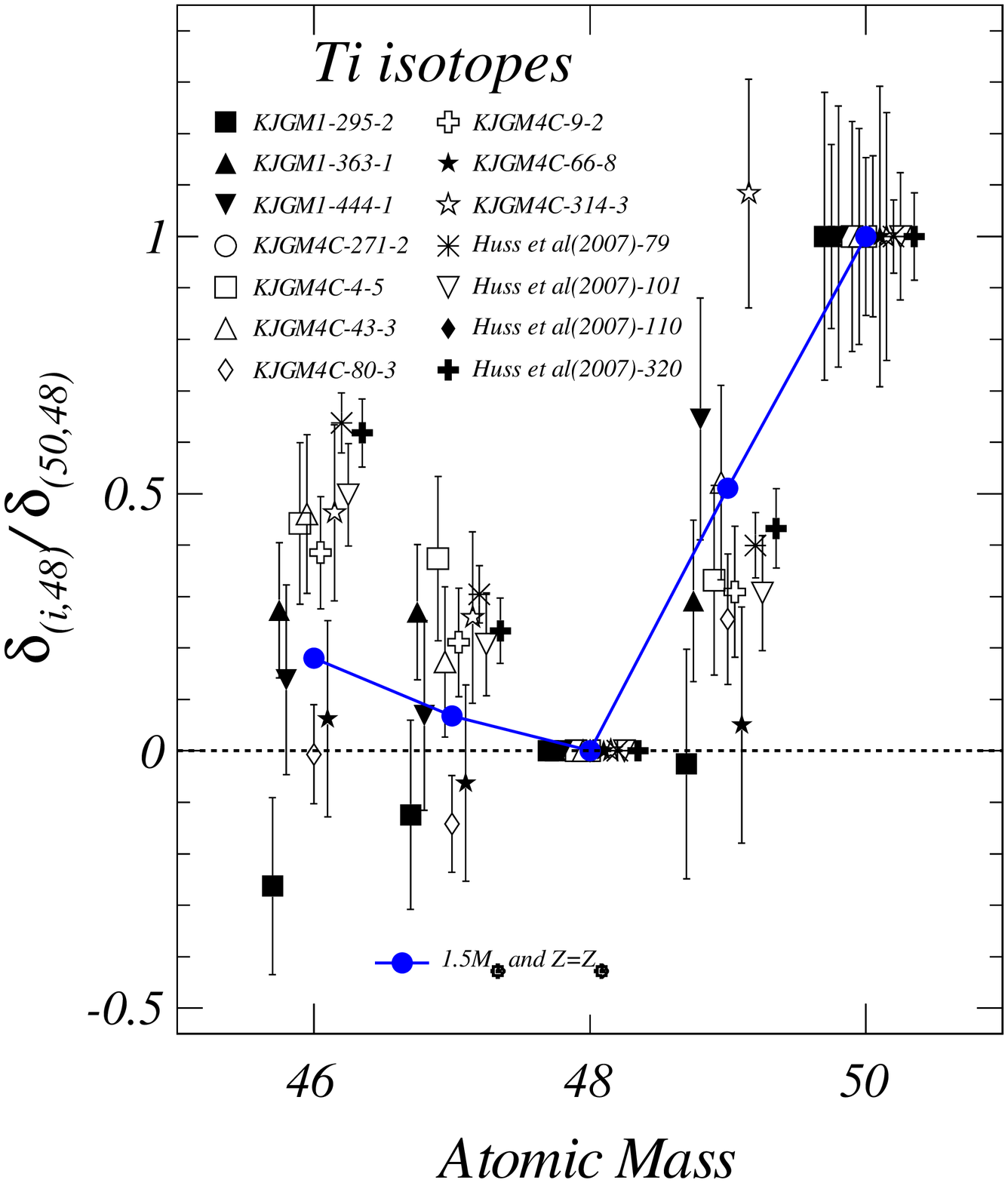}
\caption{\label{ti-2}}
\end{figure*}

\vspace{1cm}
{{\bf Figure \ref{ti-2}}. A wide selection of Ti isotopic data on main stream CIRCONS compared to AGB $s$ process patterns. These samples have $\delta_{50,46}$ down to 200 per mil. Data taken from \citet{amari1}, \citet{amari2} and \citet{huss}. For comparison we show the $Q_{i,48}$ normalized to $Q_{50,48}=1$ (that is $Q_{i,48}/Q_{50,48}$). The data shown are in the form $\left( \delta_{i,48}\pm e_{i,48}\right) /\left( \delta_{50,48}\right) $ where $e_{i,48}$ is the uncertainty quoted for that ratio. No errors were propagated from the values for $\delta_{50,48}$. It can be seen that the data cluster around the AGB model calculations, but with some clear differences. In several cases they are well within the error envelope, but there are also many exceptions. For example, sample number 320 has clear excesses of $^{46}$Ti and $^{47}$Ti relative to the model, but matches $^{49}$Ti very well.
Sample KJGM4C-314-3 appears to be quite discrepant, with a large excess of $^{49}$Ti. However, if one considers the errors in both $^{50}$Ti/$^{48}$Ti and $^{49}$Ti/$^{48}$Ti, the results could be 
compatible with AGB processing.}

\begin{figure*}
\includegraphics[width=\textwidth]{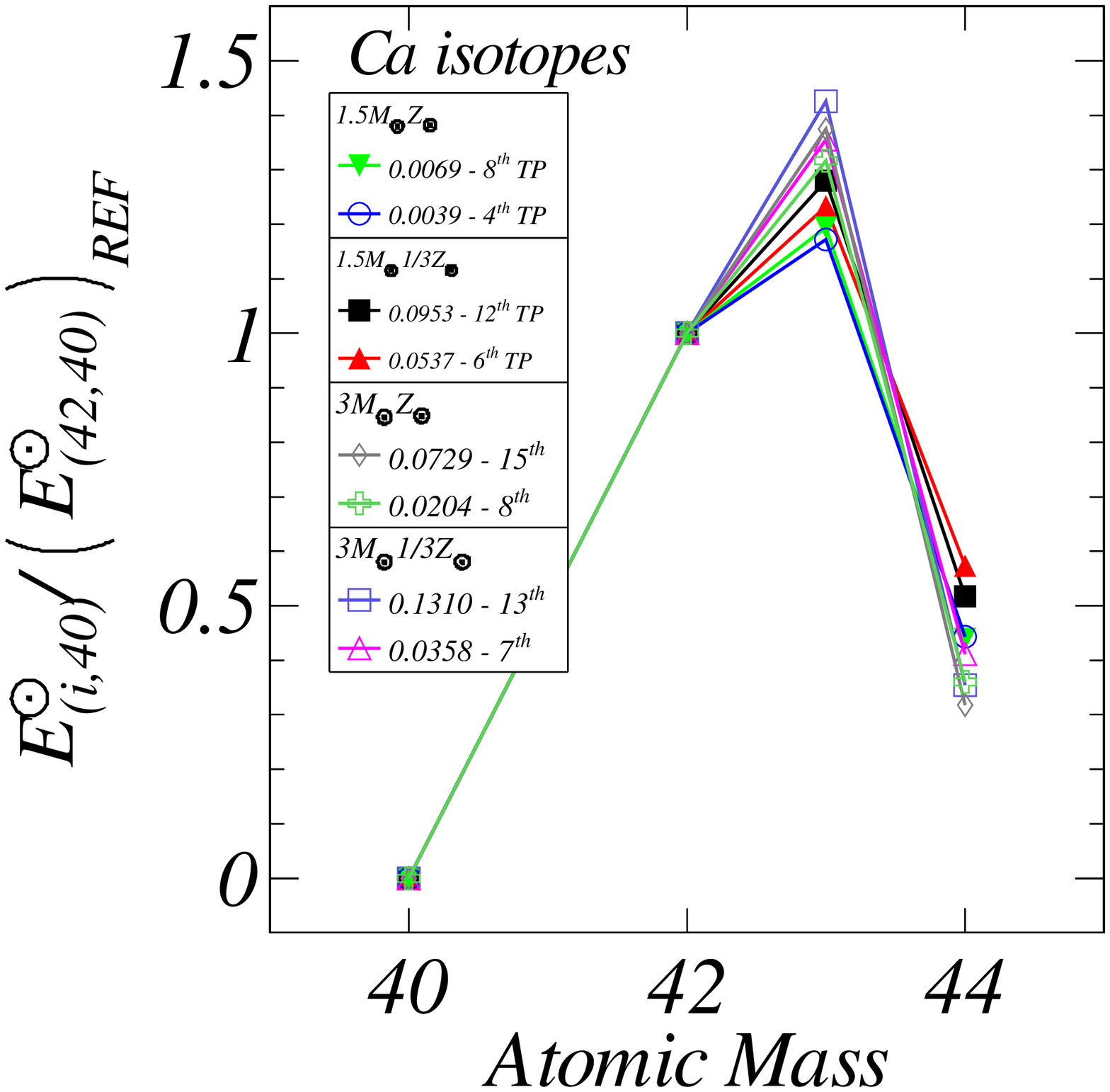}
\caption{\label{calcium}}
\end{figure*}

\vspace{1cm}
{{\bf Figure \ref{calcium}}. The enrichment factors for the Ca isotopes $^{42}$Ca, $^{43}$Ca, and $^{44}$Ca, assuming a solar initial isotopic composition. $^{46}$Ca and $^{48}$Ca are omitted. These results are normalized to $E^{\odot}_{42,40}=1$; otherwise the display is same as in Figure \ref{over}. It can be seen that $E^{\odot}_{42,40}$ and $E^{\odot}_{43,40}$ are enhanced relative to $E^{\odot}_{44,40}$ in every case. Assuming $^{42}$Ca/$^{44}$Ca as normal, then a calculation of the $^{48}$Ca/$^{44}$Ca ratio will result in an apparent excess of $^{48}$Ca for $s$-processed material; any other non-Solar source will give effects in $^{48}$Ca/$^{40}$Ca. It is seen that these results for Ca are well behaved (compare to Ca in Figure \ref{over}).}

\newpage
\newpage

\newpage

\newpage

\newpage

\newpage

\newpage

\end{document}